\documentclass[pra,showpacs,twocolumn,superscriptaddress,floatfix,aps]{revtex4-2}
\usepackage{amsmath,amssymb,bm}
\usepackage{comment}
\usepackage{color}
\usepackage{float}
\usepackage{appendix}
\usepackage{soul}
\usepackage{hyperref}
\usepackage[normalem]{ulem}
\hypersetup{colorlinks,
	linkcolor=blue,%
	citecolor=blue,%
	urlcolor=blue}
\usepackage[T1]{fontenc} 
\usepackage{graphicx}

\usepackage{bm} 
\usepackage{amsmath}
\usepackage{mathtools}
\usepackage{amssymb} 

\usepackage{array}
\usepackage{multirow}
\usepackage{color} 
\usepackage{float} 
\usepackage{graphicx} 
\usepackage{dcolumn} 
\usepackage{hyperref}
\usepackage[normalem]{ulem}
\hypersetup{colorlinks,
	linkcolor=blue,%
	citecolor=blue,%
	urlcolor=blue}
\newcommand{\mbr}{\mathbf{r}}
\newcommand{\mbx}{\mathbf{x}}
\newcommand{\mbk}{\mathbf{k}}
\newcommand{\mbu}{\mathbf{u}}

\usepackage{comment}

\begin{document}

\title[Impurities induced vortex lattice melting and turbulence in rotating BECs]{Impurities induced vortex lattice melting and turbulence in rotating Bose-Einstein condensates}
			
\author{Rony Boral}
\affiliation{Department of Physics, Indian Institute of Technology  Guwahati, Guwahati-781039, Assam, India}
	
\author{Swarup K. Sarkar}
\affiliation{Department of Physics, Indian Institute of Technology  Guwahati, Guwahati-781039, Assam, India}
\author{Paulsamy Muruganandam}
\affiliation{Department of Physics, Bharathidasan University, Tiruchirappalli 620 024, Tamil Nadu, India}

\author{Pankaj K. Mishra}%
\affiliation{Department of Physics, Indian Institute of Technology  Guwahati, Guwahati-781039, Assam, India}	
	
\date{\today}

\begin{abstract}
We investigate the impact of various impurities on rotating Bose-Einstein condensates confined within two-dimensional harmonic and optical lattice potentials. Without impurities, the rotating condensates display an organized square lattice pattern of vortices due to the influence of a square optical lattice. The introduction of impurity potentials disrupts this lattice structure, inducing a phase transition from an ordered state to a disordered state. Our analysis encompasses both static and dynamic types of impurities. The static impurities are implemented using a randomly varying potential with a spatially random amplitude. The transformation of the vortex lattice structure, in this case, relies on the strength and lattice constant of the impurity potential. For dynamical impurities, we employ a Gaussian obstacle that orbits around the condensate at a specific distance from its center. In this scenario, the vortex lattice melting occurs beyond a certain threshold radius and frequency of oscillation of the rotating obstacle. We characterize the melting of the vortex lattice due to impurities using various quantities, such as the structure factor and angular momentum. Notably, in the vortex-melted state, the angular momentum follows a power-law dependence with an exponent of approximately $1.73$, regardless of the type of impurity. Finally, we demonstrate the signature of the presence of a turbulent state within the vortex-melted state generated by both static and dynamical impurities.
\end{abstract}			
%
			%
			%
			%

\maketitle

\section{Introduction}
\label{sec1}
Since the realization of Bose-Einstein condensates (BECs) in laboratory experiments, it has become a robust platform to simulate, in a controlled way, many of the microscopic condensed matter phenomena like Anderson localization~\cite{Billy:2008, Roati:2008}, superfluid-to-mott-insulator transition~\cite{Greiner2002}, and spin-Hall effect~\cite{Leblanc2012, Beeler2013}. The other feature of the BECs is the precise generation of the topological defect manifests as a quantum vortex which in the laboratory experiment is generated using laser stirring~\cite{Inouye:2001, Raman:2001}, rotating magnetic trap~\cite{Hodby:2001, Williams1999, Williams:2010}, and phase imprinting~\cite{Leanhardt:2002, Brachmann:2011}. 

A rapidly rotating BEC, subject to the harmonic trap generates a periodic array of vortices generally arranged on the triangular or hexagonal lattices which appear to be the same as the Abrikosov lattice generated in the high $T_c$ superconductor in the presence of the strong magnetic field. As the rotating BECs are subject to the optical trap the vortices get pinned to the optical lattice and display a structural transition. For instance, several works show the transition from triangular to square, quasiperiodic, and other complex structures that solely depend upon the nature of the optical lattice~\cite{Madison:2000}. 

The melting of the vortex lattice in the presence of disorder impurities in BECs offers an intriguing insight into the interplay of rotation, interaction and disorder. The vortex lattice melting phenomenon is quite a common phenomenon in high $T_c$ superconductors, where the competition between impurity disorder and thermal fluctuations leads to the vortex melting~\cite{Rosenstein:2010, Safar:1992, Guillamon:2009}. Direct observations of the melting of two-dimensional superconducting vortex lattices have been reported in \cite{Guillamon:2009}. Despite being vastly different systems, the vortex dynamics in superfluids (BEC and liquid helium) and superconductors exhibit similar behaviour. There have been several experimental studies on vortex lattice melting in BECs. For example, Engels \cite{Engels:2003} observed the melting phenomena due to the anisotropic expansion of the vortex lattice. Coddington {\it et al.}~\cite{Coddington:2003} experimentally demonstrated that shearing force generated from laser beams destroys the vortex lattice in the condensates. The experimental observations are duly complemented by the theoretical and numerical analysis. For instance, Snoek and Stoof~\cite{Snoek:2006} extensively investigated the melting phenomenon in a one-dimensional optical lattice caused by inhomogeneous quantum fluctuations at vortex positions. Using mean-field Gross-Pitaevskii equations (GPE) several numerical investigations have been conducted to understand the dynamical behaviour of vortex lattices under the influence of various impurity types in rotating BECs. For instance, Mithun {\it et al.}~\cite{Mithun:2016} numerically analyzed the distortion of a vortex lattice in the condensate under the influence of a single impurity in the form of Gaussian potential and found that the vortex lattice makes a transition to the disordered arrangement of the vortex phase also termed as the vortex lattice melted state in presence of quasi-periodic impurity generated by the superposition of two optical lattices. Following this, melting of the Abrikosov vortex lattice structure was also reported in the presence of random impurity~\cite{Mithun:2018, Hu:2020}.

On the other hand in recent years there has been a significant surge in understanding the turbulence in superfluid, particularly through the understanding of the dynamics of vortices in BECs~\cite{tsubota:2009}.  Several experimental~\cite{Henn:2009, Gauthier:2019,navon:2019} and numerical~\cite{Kobayashi:2005, kobayashi:2007, kobayashi:2008} studies have been performed on the turbulent behaviour of BECs. In a two-dimensional (2D) fluid, the existence of the inverse energy cascade~\cite{Kraichnan:1967, Kraichnan:1975}, where energy flows from smaller to larger spatial length scales, plays a crucial role in understanding turbulent behaviour. Remarkably, the inverse energy cascade of turbulent flows is effectively described by the well-known Kolmogorov's $k^{-5/3}$ scaling law in the incompressible kinetic energy spectrum, which has been firmly established in this field~\cite{kolmogorov:1997, Kobayashi:2005}. Several techniques have been used to generate the turbulence in BECs. Reeves {\it et al.}~\cite{Reeves:2012} numerically analyzed the turbulent BECs where the turbulence is generated via stirring the condensate with a Gaussian obstacle barrier. They identified three distinct regions related to the vortex dynamics and associated turbulent behaviour by analyzing the power law exponent of the kinetic energy spectrum. Neely {\it et al.}~\cite{Neely:2010} used the Gaussian obstacle to observe vortex dipoles. Later on, Mithun {\it et al}~\cite{Mithun:2021} used the Gaussian obstacle potential to generate turbulence in BECs. In contrast, their study demonstrated the decay of turbulent behaviour by tuning the intercomponent interaction strength.  Das {\it et al.}~\cite{Subrata:2022} has rigorously explored the vortex dynamics by incorporating the rotating paddle potential~\cite{white:2012} in a two-species binary condensate and reported the vortex clustering due to the counter-rotating of two paddle potentials. Recently, Anirudh {\it et al.} studied turbulence in rotating BECs and BEC mergers~\cite{Sivakumar2024, Sivakumar2024a}. Furthermore, the incompressible spectrum exhibited power-law behaviour with exponents of $k^{-5/3}$ and $k^{-3}$ for low and high wavenumbers, respectively. In a subsequent study, Silva {\it et al.}~\cite{silva:2023} investigated the turbulent behaviour of a mass-imbalanced binary BEC by perturbing the harmonic trap using a time-dependent periodic potential. This approach is analogous to the stirring of the condensate, which leads to turbulent dynamics in the system.

So far, the turbulent behaviour in a condensate has been investigated for the condensate stirring through different approaches either without the trap or with the trap under harmonic potentials. However, the analysis of turbulence in the vortex lattice melting state, where the organized arrangement of vortices becomes completely disordered due to strong inter-vortex interactions, is still lacking.  In this paper, we utilize a slightly different protocol to analyze the effect of various types of impurity potential on the square lattice pattern of vortices. In the absence of impurities, the rotating condensates exhibit a square lattice structure due to the pinning of the vortices on square optical lattice potential. As impurity is introduced the vortex lattice makes a transition from the ordered to a disordered state. In our analysis, we consider both dynamical and static types of impurities, specifically a Gaussian obstacle and random static impurity, respectively. In the case of static impurity, the randomness is generated by varying the strength of the impurity, whereas for dynamic impurities, the Gaussian obstacle is stirred throughout the condensate by revolving it circularly. In both scenarios, the transition from the ordered vortex lattice structure to the melted vortex lattice structure has been systematically captured by examining various quantities such as structure factor, angular momentum, information entropy, and different energies of the condensate. One of the main aims of this work is to explore the turbulent feature of the melted vortex lattice state in the presence of both static as well as dynamic impurity potential. By analyzing the kinetic energy spectrum, we establish the presence of the turbulent nature of the condensate in the vortex lattice melted state in the case of both the impurities that indicate the universality of the melted vortex lattice state irrespective of the mechanism through which it has been generated. 
  
The paper is structured as follows. In section~\ref{sec:2}, we provide the governing equations and details of different static and dynamical potentials used in the paper to trigger the melting of the vortex lattices. Section~\ref{sec:3} presents a brief description of the simulation details and several quantities used to characterize the different states of the vortex lattices. The description includes the definitions of structure factor, energy, angular momentum, and drag force associated to the condensate density. In section~\ref{sec:4}, we present the numerical simulation results in the case of random impurity potential and Gaussian obstacle potential. For each type of potential, we investigate the impact of various parameters, such as impurity strength or impurity lattice constant for random static impurity, and further, the position or oscillation frequency of the Gaussian obstacle on the vortex lattice structure. In section~\ref{conclusion}, we finally conclude our work.

\section{Mean-field model}
\label{sec:2}

We consider a BEC confined in a transverse trap that rotates with an angular velocity $\Omega$ modelled using the following dissipative Gross-Pitaevskii equation
(GPE)~\cite{Kasmatsu:2006}:
\begin{align}
(i-\gamma)\frac{\partial\psi(x, y, t)}{\partial t}= &\Bigg[-\frac{1}{2}\nabla^2+V_{\rm ext}(x, y)+g\left\vert\psi(x,y, t)\right\vert^{2} \notag \\ &
-\mu-\Omega L_{z}\Bigg]\psi(x,y,t),
\label{eqn:1}
\end{align} 
where $\nabla^2 \equiv \partial_x^2 + \partial_y^2$ is the Laplacian operator, $V_{\rm ext}(x, y)$ is the external trapping potential, $g= 4 N \sqrt{\pi \lambda} a_s/a_{\perp}$ is the nonlinear interaction with $a_{s}$ being the $s$-wave scattering length and $N$, the total number of atoms in the condensate and $\lambda= \omega_z/\omega_{\perp}$ is the ratio of longitudinal trap frequency $\omega_z$ and transverse trapping frequency $\omega_{\perp}$. $L_{z}=-i\hbar(x\frac{\partial}{\partial y}-y\frac{\partial}{\partial x})$ is the angular momentum operator and $\mu$ is the chemical potential given as
\begin{align}
\mu = \int \Bigg[\frac{1}{2}\left\vert \nabla \psi \right\vert^{2}+V_{\rm ext}\left\vert\psi\right\vert^{2}+ g\left\vert\psi\right\vert^{4}-\Omega\psi^{\dagger}L_z\psi \Bigg] dx dy \label{eqn:mu}
\end{align} 
The dimensionless equation (\ref{eqn:1}) has been obtained by choosing $a_{\perp} = \sqrt{\hbar/m\omega_{\perp}}$ (with $m$ is the mass of an atom) as the characteristic length scale, $\omega^{-1}_{\perp}$ as the characteristic time scale, and $\hbar\omega_{\perp}$ as the energy scale. The wave-function is rescaled as $\psi(x,y,t) = \tilde{\psi}(x,y,t) a_{\perp}^{3/2}/ \sqrt{N}$, where $\tilde{\psi}(x,y,t)$ is the dimensionless wave function. For brevity, we omit the tilde sign over the dimensionless wave function. A dissipation term, denoted by the symbol $\gamma$, has been incorporated to account for the effects of temperature and the non-condensate fraction~\cite{Jin:1997, Marago:2001}. We consider $\gamma=0.03$~\cite{Kasmatsu:2006, Choi:1998, Kasamatsu:2002, Kasamatsu:2003, kasamatsu:2011}.



For our analysis we consider the external potential as, 
\begin{align}
V_{\rm ext}=\frac{1}{2}\left( x^{2} + y^{2}\right)+V_{\rm OL}+V_{\rm imp},
	\label{eqn:2}
\end{align}
where $V_{\rm OL}$ represents the optical lattice potential which can be realized in experiments through the superposition of four counter-propagating laser beams with wavelength $\hat{\lambda} \approx 830\,$nm~\cite{Williams:2010} that consider the form as~\cite{Mithun:2014},
\begin{align}
V_{\rm OL}(\bm{r})=\sum_{n_1}\sum_{n_2} V_0 \exp\left[-\frac{ \lvert\bm{r}-\bm{r}_{n_1, n_2}\rvert^{2}}{(\sigma/2)^2} \right],
\label{eqn:3}
\end{align}
where $\bm{r}_{n_1, n_2} = n_1 \bm{a_1} + n_2 \bm{a_2}$ denote the lattice vector with $n_1$ and $n_2$ being the number of lattice sites along the $x$- and $y$- direction respectively. $V_0$ is the strength of the optical lattice potential, and $\sigma$ represents the width of the Gaussian laser beam~\cite{Fort:2005}. For our analysis, we have considered the lattice unit vector as $\bm{a_1} = a\hat{i}$, $\bm{a_2} = a\hat{j}$ to obtain the square lattice, where $a$ represents the lattice constant. In the present, we vary both the lattice constant as well as the potential strength ($V_0$) to obtain the vortex lattice state. Throughout the study, the Gaussian beam waist of the laser is kept at $\sigma = 0.65$. 

The main aim of the work is to investigate the effect of the static and dynamic impurities. In the first part, we focus on the static impurity with randomly distributed amplitudes in 2D, which in the experiment can be realized through the laser speckle method~\cite{Sanchez:2010}. The random impurity potential used in our work has the form given by~\cite{Mithun:2016}:
\begin{align}
V_{\rm imp} =\sum_{n_1}\sum_{n_2} V_0^{\rm imp} \left[-\frac{ \lvert\bm{r}-\bm{r}_{n_1, n_2}^{\rm imp}\rvert^{2}}{(\sigma/2)^2} \right],
	\label{eqn:6}
\end{align}
where $\bm{r}_{n_1, n_2}^{\rm imp}=n_1 \bm{a_{1}^{\rm imp}}+n_2 \bm{a_{2}^{\rm imp}}$ denotes the lattice vector of impurity potential with $n_1$ and $n_2$ being the number of lattice sites along the $x$- and $y$- direction, respectively. We consider the square geometry of the random potential with two lattice unit vectors given by $\bm{a_{1}^{\rm imp}} = a_{\rm imp}\hat{i}$, $\bm{a_{2}^{\rm imp}} = a_{\rm imp}\hat{j}$, where $a_{\rm imp}$ is the lattice constant corresponding to the impurity potential. $V_0^{\rm imp}$ is the strength of the impurity potential. Here, the amplitude of the potential at different lattice sites is randomly distributed whose value lies in the range of $[-V_0^{\rm imp}, V_0^{\rm imp}]$. In the second case, we will investigate the impact of dynamically generated impurities, which can be realized through stirring a laser beam in the condensate~\cite{Gauthier:2019, Neely:2013, Johnstone:2019, Kwon:2014}. 

The single Gaussian obstacle is defined as,
\begin{align}
V_{\rm obstacle}= V_{0}^{\rm obs} \exp\left[-\frac{ (x-x_{0}(t))^{2}+(y-y_0(t))^{2}}{d^{2}}\right],
\label{eqn:5} 
\end{align}
where $x_0 = r_0\cos(\omega t)$ and $y_0 = - r_0 \sin(\omega t)$ are the $x$- and $y$-coordinates of the obstacle which have oscillation frequency $\omega$ with $r_0$ is the distance of the obstacle from the trap center. The amplitude of the obstacle is set to be $V_{0}^{\rm obs} = 80\hbar\omega_{\perp}$. We consider the size of the obstacle $d=0.3$ for all the simulation runs.

\section{Numerical details and different quantities to characterize the structure and dynamics of vortex lattice}
\label{sec:3}

We use the GPE-lab software that uses the implicit-time splitting pseudo spectral method~\cite{Antoine:2014} to solve the governing mean-field GPE (\ref{eqn:1}) with a space step $\Delta x= \Delta y=0.1860$ and time step $\Delta t=0.001$. To make our simulation results viable to the experiment, we choose $N=2\times10^4$ atoms corresponding to the condensate of $\rm ^{87}Rb$ atom. The axial and transverse frequency of the harmonic trap is considered as $\omega_z=2\pi \times 100$ Hz and $\omega_{\perp}=2\pi \times 10$ Hz, respectively. Considering the scattering length of $\rm ^{87}Rb$ as $a_s=5.5$nm~\cite{Kasamatsu:2003, Madison:2001} with the characteristic length scale as $a_{\perp}=2.41\mu$m. We fix the other parameters as $g=1000$ and $\gamma= 0.03$.
In order to conserve the norm of the condensate wave function for a nonzero value of $\gamma$, the chemical potential has been adjusted at each time step. In our simulation, the chemical potential at each step is corrected using the following expression~\cite{Mithun:2018}: 
\begin{align}
	\Delta \mu = (\Delta t)^{-1} \ln \frac{\int dx dy \lvert\psi(x,y,t)\rvert^{2}}{\int dx dy\lvert \psi(x,y, t+\Delta t) \rvert^{2}}, \notag 
\end{align}


In this work, we utilize the structure factor to characterize the structural transformation of the vortex lattice of the rotating BECs. The structure factor is defined as $S(\bm{k})=\frac{1}{N_c}\sum_{i=1}^{N_v}n_i \exp ({i\bm{k}\cdot \bm{r_i}})$, where $n_i$, $\bm{r_i}$, and $N_c$ denote the winding number, position of the $i$th vortex, and the total winding number, respectively. $N_v$ is the total number of vortices present in the condensate. The nature of the vortex lattice is determined based on the value of $S(k)=\lvert S(\bm{k})\rvert$. For instance, with a perfect square lattice, $S(k)$ lies between $0.5 \lesssim S(k) \lesssim 1.0$. For the situation when $S(k) < 0.5$, we characterize the state of the vortex lattice as deviated from the square structure and thus termed as in the melted state. 


To elucidate the transition from the vortex lattice state to the melting state, we have also utilised the total energy of the condensate as one of the relevant physical quantities. The total energy of the system is defined as~\cite{Antoine:2014}, 
\begin{align}
	E_T = \int \Big[\frac{1}{2}\left\vert\nabla \psi \right\vert^{2}+V_{\rm ext}\left\vert\psi\right\vert^{2}+\frac{g}{2}\left\vert\psi\right\vert^{4}-\Omega\psi^{*}L_z\psi \Big] dx dy,
	\label{Eq_Energy}
\end{align}
where the first, second, third, and fourth terms represent the kinetic ($E_K$), the potential ($E_{pot}$), the interaction ($E_{int}$), and the rotational energy ($E_{rot}$) of the condensate, respectively.


To characterize the melting state of the vortex lattice we compute the energy spectrum and show the presence of the turbulence-like features of the condensate. The kinetic energy spectra have two components, namely compressible and incompressible parts in which the former is related to the sound wave production and later is related to the vortices~\cite{kolmogorov:1997, Subrata:2022, Angela:2014}. The scaling laws in the kinetic energy spectrum provide us with a deeper understanding of the turbulent behaviour in a fluid~\cite{Kobayashi:2005, Reeves:2012, Madeira:2020}. The kinetic-energy $\frac{1}{2}\left\vert \nabla \psi \right\vert^{2}$ in Eq.~(\ref{Eq_Energy}) can be written as:
\begin{align}
\frac{1}{2}\vert\nabla\psi\vert^2= \frac{1}{2} \left( \rho(\bm{r})\left\lvert\bm{u}\right\rvert^{2}+ \Bigl\lvert \nabla \sqrt{\rho(\bm{r})} \Bigr\rvert^{2} \right),
\end{align} 
where the Madelung transformation $\psi(\bm{r})=\sqrt{\rho(\bm{r})} e^{\mathrm{i} \phi}$ yields the condensate density and the superfluid velocity $\bm{u}=\bm{\nabla} \phi$. The first and second terms represent, respectively, the kinetic energy density $E_{K}$ and the quantum pressure $E^{q}$. The different component of energy are given by 
\begin{align}
E_{K} & =\frac{1}{2}\int \rho(\bm{r})\left\vert\bm{u}\right\vert^2 d^2 \bm{r} \\
E^q & =\frac{1}{2}\int \left\vert\nabla \sqrt{\rho(\bm{r})}\right\vert^2 d^2 \bm{r}. 
\end{align}
Here, the velocity $\bm{u}$ is decomposed as a sum of the incompressible $\bm{u}_{ic}$ and compressible $\bm{u}_c$, such as $\bm{u} = \bm{u}_{ic} + \bm{u}_c$, where $\nabla \cdot \bm{u}_{ic} = 0$ represents the solenoidal part and $\nabla \times \bm{u}_c = 0$ represents the irrotational part.
Following Parseval's theorem, different energy components can be represented as
\begin{align}
E^{\zeta}_{\mathrm{kin}} = \frac{m}{2}\int d^2\mbk\lvert\tilde{\mbu}^{\zeta}(\mbk)\rvert^2,
\label{eq:parseval}
\end{align}
where $\zeta \in \{i,c,q\}$ being incompressible, compressible and quantum pressure components of the kinetic energy, respectively and 
\begin{align}
\tilde{\mbu}^{\zeta}(\mbk) = \frac{1}{2\pi}\int d^2\mbr e ^{-i\mbk\cdot\mbr}\mbu^{\zeta}(\mbr)
\label{eq:ftransform}
\end{align}
Transforming equation (\ref{eq:parseval}) into the cylindrical coordinates in Fourier space ($k$ space) for a 2D condensate, we obtain
\begin{align}
E_{\mathrm{kin}}^{\zeta} = \int_0^{\infty} dk \varepsilon_{\mathrm{kin}}^{\zeta}(k) .
\label{eq:ekinzeta}
\end{align}
In general, to compute the energy spectra $\varepsilon_{\mathrm{kin}}^{\zeta}$, data are first  binned in the wavenumber ($k$) space, and then summed over angular intervals. This approach is suitable for smalll-$k$ regimes. However, to obtain a more precise spectrum, we have used the analytical evaluation of $k$-space integrals and its numerical implementation developed by Bradley and coworkers~\cite{Bradley2022}. The scheme utilizes the angle-averaged Wiener-Khinchin theorem, which connects the spectral densities with an associated correlation function.
Following this, $\varepsilon_{\mathrm{kin}}^{\zeta}$ from equation (\ref{eq:ekinzeta}) could be represented as~\cite{Bradley2022}
\begin{align}
\varepsilon_{\mathrm{kin}}^{\zeta}(k) = \frac{m}{2}\int d^2\mbx\Lambda_2(k,\lvert\mbx\rvert)C \left[\mbu^{\zeta}, \mbu^{\zeta}\right](\mbx),
\label{eq:specden}
\end{align}
where $\Lambda_2(k,\lvert \mbx \rvert) = (1/2\pi) k J_0(k\lvert\mbx\rvert)$ is the 2D kernel function, involving the Bessel function $J_0$ and $C[\mbu^{\zeta}, \mbu^{\zeta}](\mbx)$ represents the two-point auto-correlation function in position for a given velocity field. The above relation implies that for any of the position-space fields $\mbu^{\zeta}$, there exists a spectral density [see equation (\ref{eq:specden})], that is, equivalent to an angle-averaged two-point correlation in $k$ space.

To have a comprehensive understanding of the transition from an ordered vortex lattice to the melted-vortex-lattice state, we have used the total information entropy ($S_{\rm enp}$) for both real space and momentum space. In real space, the entropy ($S_r$) can be expressed as
\begin{align}
	S_{r}=-\int \rho(\bm r)\log \rho(\bm r) d^{2}\bm r
\end{align}
Where, $\rho(\bm r) = \left\vert\psi(\bm r)\right\vert^2$ represents probability density in the real space. The entropy in momentum space $S_k$ is given by
\begin{align}
	S_{k}=-\int \rho(\bm k)\log \rho(\bm k) d\bm k,
\end{align}  
where $\rho(\bm k)=\left\vert\psi_k\right\vert^2$ is the probability density in the Fourier space. The total entropy  $S_{\rm en} = S_r + S_k$ has the contribution of both real space and momentum space entropy~\cite{Kumar:2019}. The finite value of the entropy characterizes the extent of disorderness present in the system. For our analysis, the $S_{\rm en}$ is rescaled with respect to the entropy corresponding to condensate without having an impurity ($S_{\rm en}^0$).
\begin{figure*}[!ht]
\centering
\includegraphics[width=\linewidth]{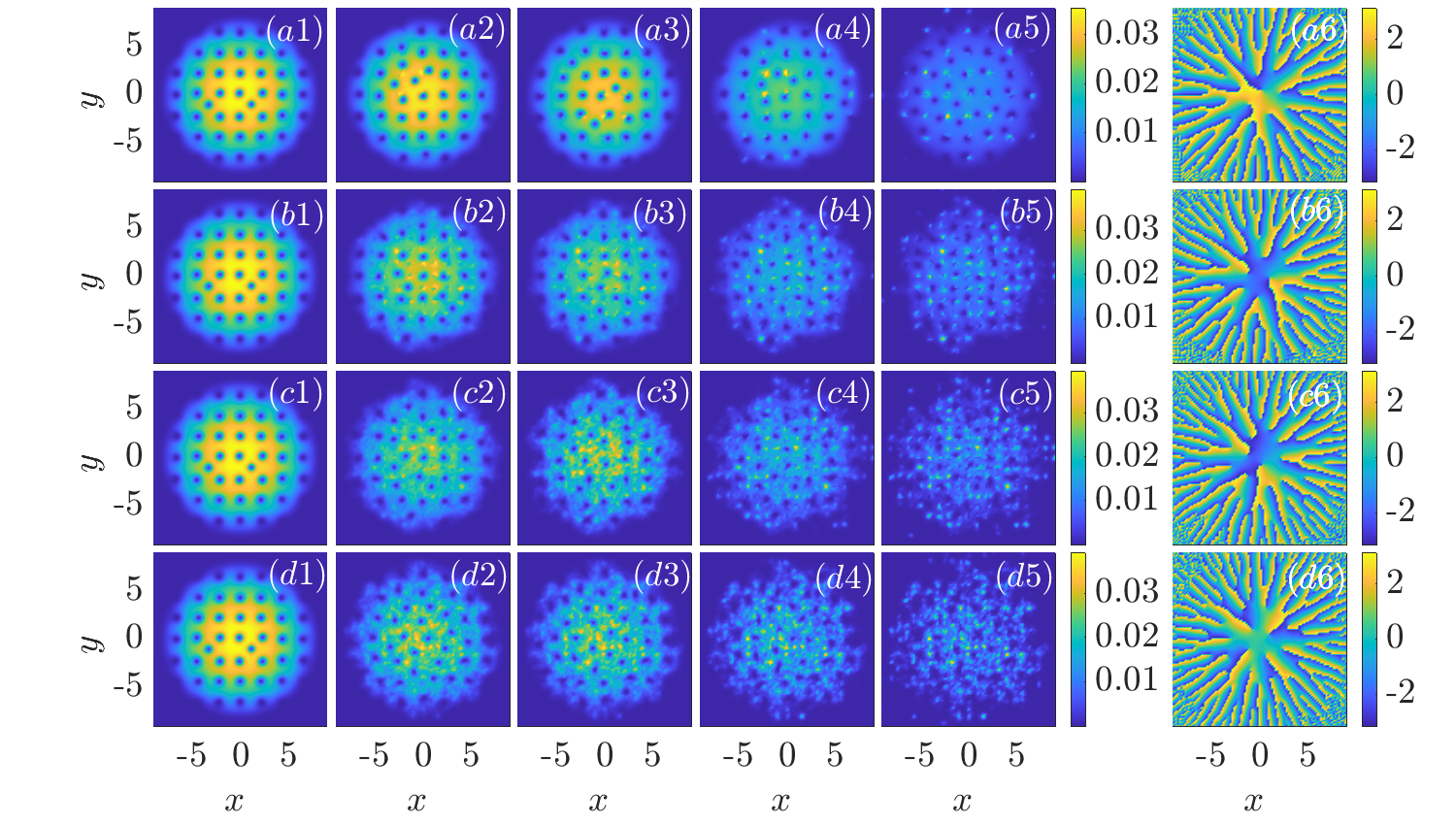}
\caption{Pseudocolor representation of condensate density in real space for different $a_{\rm imp}$ and $\eta=V_0^{\rm imp}/{V_0}$. (a1)-(a5) for fixed $a_{\rm imp}=2.2$ and $\eta$ = ($0, 3.2, 7.6, 9.2, 12$); (b1)-(b5): $a_{\rm imp}=1.3$, $\eta  = (0, 3.2, 7.6, 9.2,12$); (c1)-(c5) at $a_{\rm imp}=1$, $\eta$ = ($0, 3.2, 7.6, 9.2,12$); and (d1-d5) at $a_{\rm imp}=0.8$, $\eta=(0, 3.2, 7.6, 9.2,12$). The other parameters are $a=2.2$, $\Omega=0.8$, $V_0=5$ and $g = 1000$. Last column (a6, b6, c6, d6) represents the phase of the condensate corresponding to the vortex melting state (a5, b5, c5, d5) respectively.
}
\label{fig:real-den-static}
\end{figure*} 

\section{Numerical Results}
\label{sec:4}
In this section, we present the analysis of the structural transition of the vortex lattice from the triangular lattice to the square lattice structure upon tuning the strength of the optical lattice. We have considered the effect of two kinds of impurities on the vortex lattice structure of the condensate. First is the random Gaussian potential, for which we investigate the impact of this on the vortex lattice structure by varying its lattice constant and its strength. In the second case, we consider the oscillating Gaussian obstacle and explore its effect on the lattice structure of the vortex. 




In the next section, we present the effect of the random static impurity Gaussian impurity on the vortex lattice structure, followed by the impact of the oscillating Gaussian obstacle on the vortex lattice structure. 

\subsection{Vortex lattice melting in the presence of random static impurity}
\label{sec:vortexMelt}

We examine the influence of a random static impurity on the vortex lattice structure of a condensate. The details of the random impurity potential (\ref{eqn:6}) are given in Sec.~\ref{sec:2}. We begin the analysis of the repercussions of the impurity potential in two ways. Firstly, by changing the impurity lattice constant ($a_{\rm imp}$) and secondly, by increasing the strength of the random potential ($V_0^{\rm imp}$). Following this, we characterize the structural transformation induced by the impurity. At the end, we compute the kinetic energy spectra to investigate the turbulent behaviour of the condensate. In our study, we introduce a dimensionless ratio $\eta = V_0^{\rm imp}/ V_0$ act as the impurity strength, to scale the strength of the disorder with respect to the amplitude of the optical lattice. 
\begin{figure*}[!ht]
\centering
\includegraphics[width=0.8\linewidth]{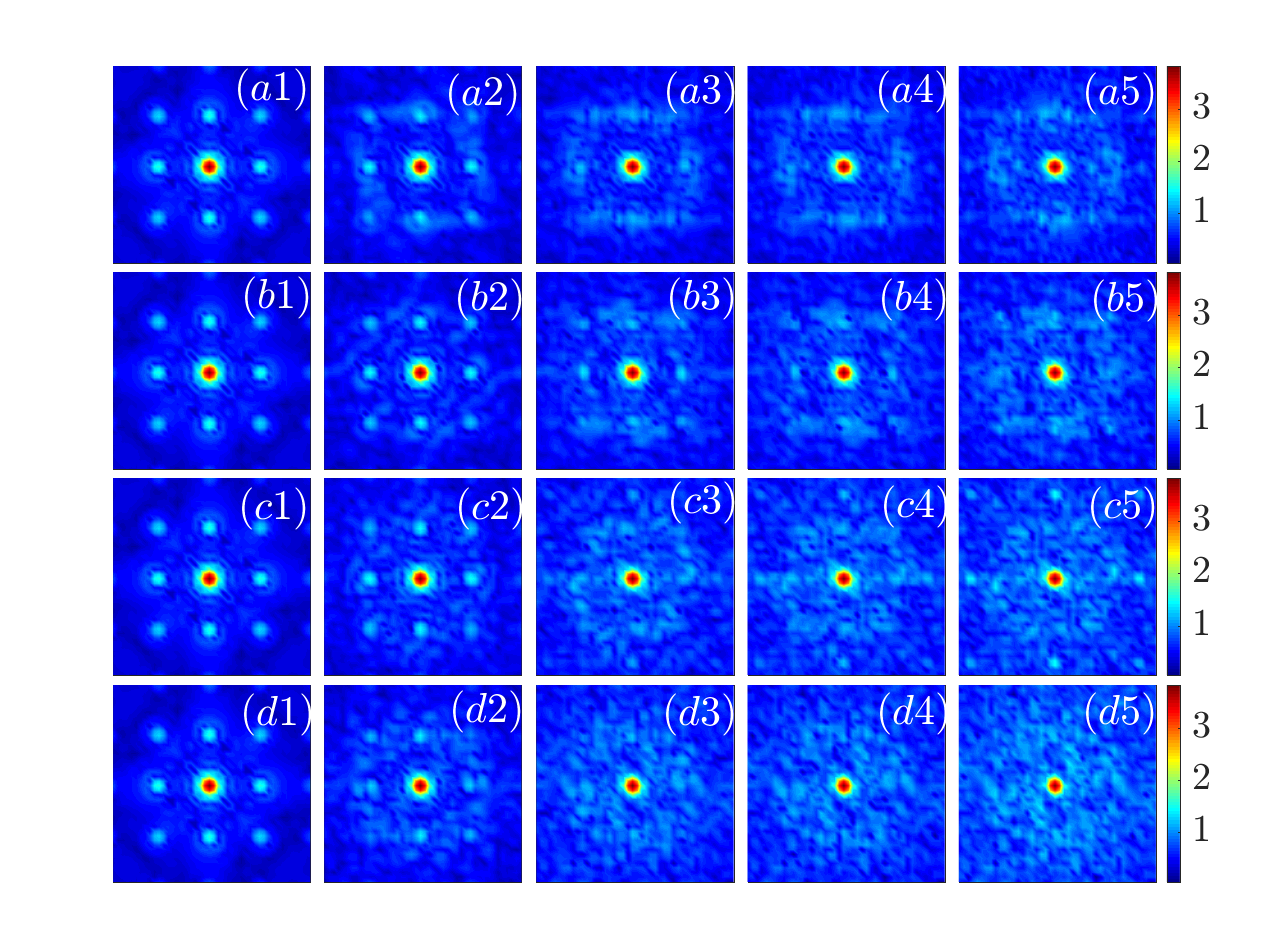}
\caption{Pseudocolor representation of the condensate density in momentum space: (a1)-(a5) for fixed $a_{\rm imp}=2.2$ and $\eta$ = ($0, 3.2, 7.6, 9.2, 12$); (b1)-(b5): For fixed $a_{\rm imp}=1.3$ and $\eta  = (0, 3.2, 7.6, 9.2,12$); (c1)-(c5) at $a_{\rm imp}=1$ and $\eta$ = ($0, 3.2, 7.6, 9.2,12$); and (d1-d5) at $a_{\rm imp}=0.8$ and $\eta$=($0, 3.2, 7.6, 9.2,12$). The other parameters are the same as figure \ref{fig:real-den-static}.}
\label{fig:kspace-den-static}
\end{figure*} 

\begin{figure*}[!ht]
\centering
\includegraphics[width=0.8\linewidth]{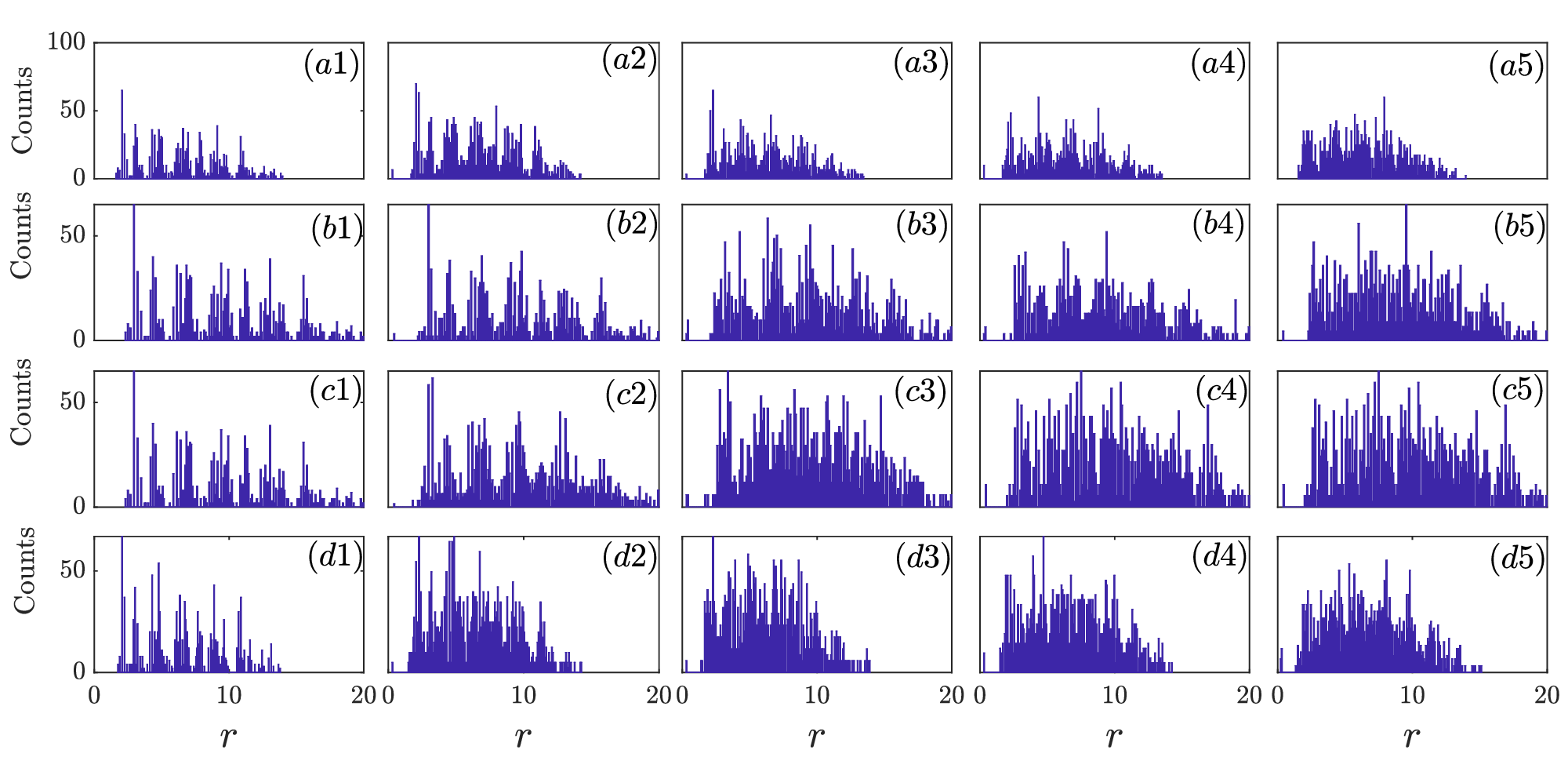}
\caption{Histogram of separation between the the pair of vortices ($r$) for different $a_{\rm imp}$ and $\eta$. The variation of $a_{\rm imp}$ and $\eta$ along rows and columns are the same as that of figure \ref{fig:real-den-static}. The presence of well-separated peaks indicates the presence of long-range order for the vortex lattice. High $\eta$ shows the presence of the vortices at all separations, indicating the melting of the vortex lattice.}
\label{fig:histogram}
\end{figure*} 
\begin{figure}[!ht]
\includegraphics[width=\linewidth]{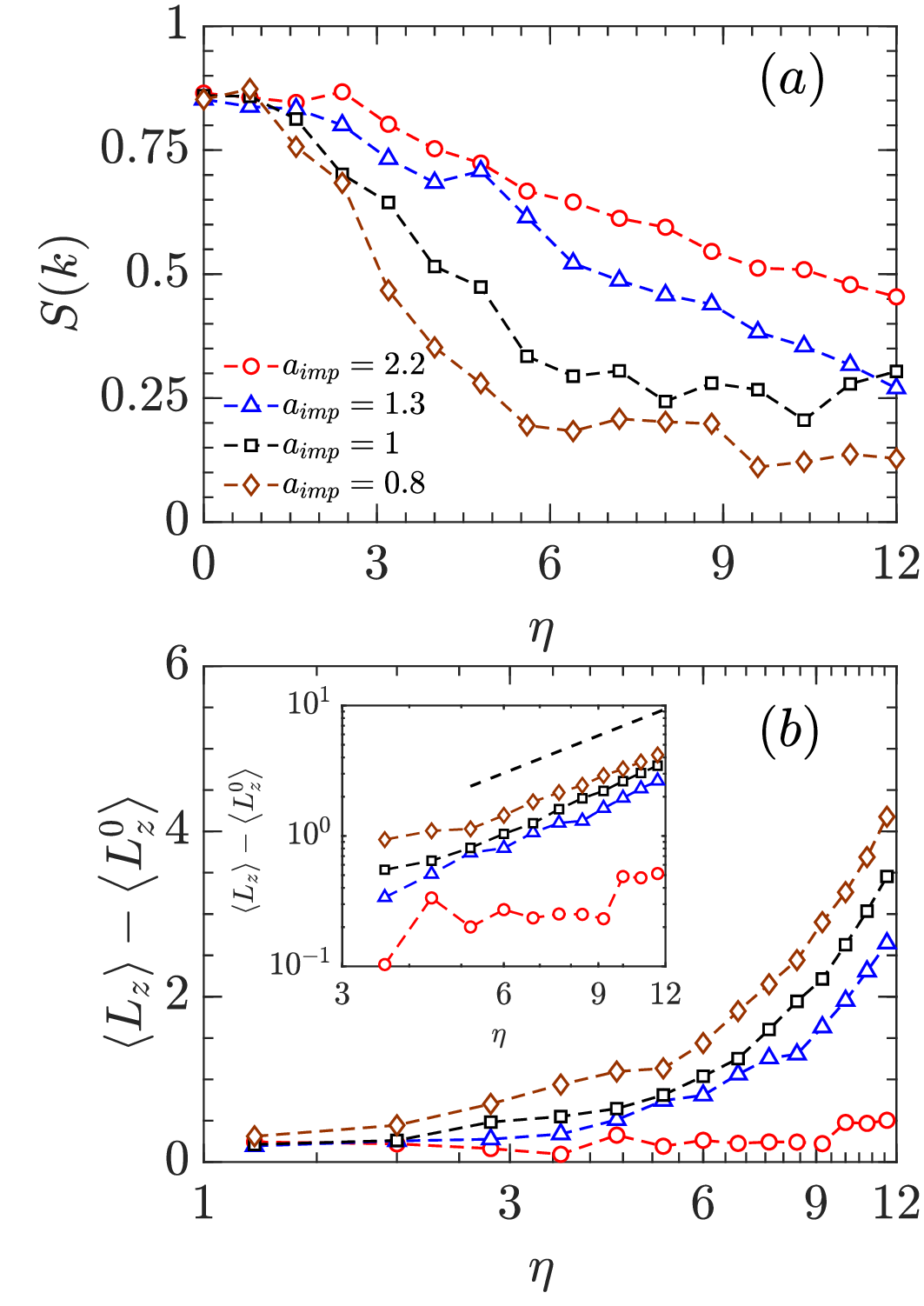}
\caption{(a) Variation of structure factor ($S(k)$) and (b) angular momentum ($ \left\langle L_{z} \right\rangle -\left\langle L^0_z\right \rangle$) as a function of $\eta$ for different impurity lattice constants ($a_{\rm imp}$). The inset in (b) shows the variation $\left\langle L_{z} \right\rangle -\left\langle L^0_z\right\rangle$ in the log scale. Here, $\left\langle L^0_z \right\rangle$ is the angular momentum of the initially pinned vortex lattice. The black dashed line shows the power law behaviour at exponent $\sim 1.73$ with $\eta$ for low $a_{\rm imp}$.
}
\label{fig:ang-strct}
\end{figure} 

\begin{figure}[!ht]
\centering
\includegraphics[width=0.95\linewidth]{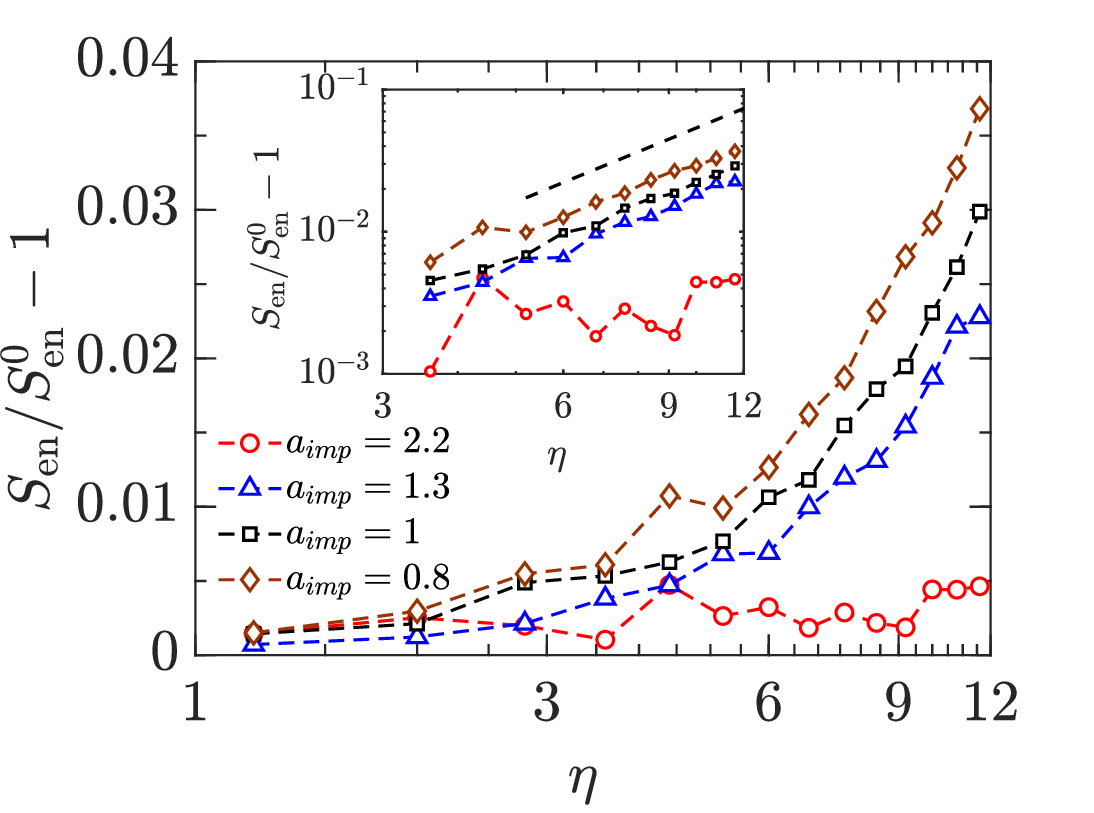}
\caption{Variation of information entropy ($S_{\rm en}$) with respect to $\eta$ for different values of $a_{\rm imp}$. The inset represents the $S_{\rm en}$ in logscale, which depicts a power-law behaviour of $S_{en}$ for larger $\eta$ with an exponent of $1.73$. The black dashed line guides the eyes for the power-law fitting. Note that, $S_{\rm en}$ rescaled with entropy of the pinned lattice, $S_{\rm en}^0$ at $V_0 = 5.0$.}
\label{fig:entropy-static}
 \end{figure}
In figure \ref{fig:real-den-static}, we show the pseudo color density profile at different $\eta$ and $a_{\rm imp}= 2.2, 1.3, 1$, and $0.8$. We find that for $a_{\rm imp}=2.2$, the vortex lattice structure remains square even at high impurity strength~($\eta$) as shown in figure \ref{fig:real-den-static}(a1)-(a5). However, as we decrease the impurity lattice constant of the random potential to $a_{\rm imp}=1.3$, the vortex lattice begins to distort from the square pattern at $\eta \sim 7.6$ [see figure \ref{fig:real-den-static}(b3)]. The squareness of the vortex lattice is completely destroyed upon further increase of $\eta$ ( see figure~\ref{fig:real-den-static}(b4, b5)). We observe that at $a_{\rm imp} = 1$, obliteration of the lattice structure starts around $\eta=3.2$~[see figure \ref{fig:real-den-static}(c2)]. It implies that the critical value of $\eta$ required to induce the lattice deformation decreases with $a_{\rm imp}$, which is quite evident from figure \ref{fig:real-den-static}(d1)-(d5). The last column of the figure shows the phase of the condensate corresponding to the column (a6-d6). The phase clearly shows the presence of disordered arrays of the vortices in the melted vortex lattice state.   

To confirm the periodic nature of the lattice, we also compute density in Fourier space as shown in the figure~\ref{fig:kspace-den-static}. The first column illustrates that for $\eta=0$, the periodic peaks form a square pattern. The peaks become more irregular for the higher values of $\eta$, which reveals the melting of the vortex lattice structure. To understand the transition from the lattice vortex state to the melted vortex lattice, in figure~\ref{fig:histogram}, we show the histogram calculated for different separations of the vortices. The histogram exhibits well-separated peaks in the vortex lattice state for small $\eta$. Upon increase in $\eta$ (moving along the row), the peaks become less distinguishable, indicating the destruction of long-range order and revealing the melting of the vortex lattice structure. This can be observed in the first column of figure~\ref{fig:histogram}~(a1)-(d1), where the peaks are well separated, compared to the subsequent columns where the separation becomes less pronounced.

Further, to quantitatively analyze the structural changes in the condensate, we examine the variation of the structure factor $S(k)$ with $\eta$ for different impurity lattice constant $a_{\rm imp}$ as depicted in figure~\ref{fig:ang-strct}(a). We observe a steady decrease in $S(k)$ as $\eta$ increases and the critical value of impurity strength~($\eta_c$), at which $S(k)$ falls below $0.5$ depends on $a_{\rm imp}$. Note that, here, we define the range of $0.5 \lesssim S(k) \lesssim 1.0$ to represent the pinned lattice state, while $S(k) \lesssim 0.5$ indicates the melted state~\cite{Pu:2005}. Specifically, at $a_{\rm imp}=1.3$ the vortex lattice structure exhibits melting behavior beyond $\eta = 8.0$, while for $a_{\rm imp}=1.0,$ and $0.8$, the melting behavior commences above $\eta=5.2$ and $4.0$, respectively. 
\begin{figure}[!ht]
\centering
\includegraphics[width=0.95\linewidth]{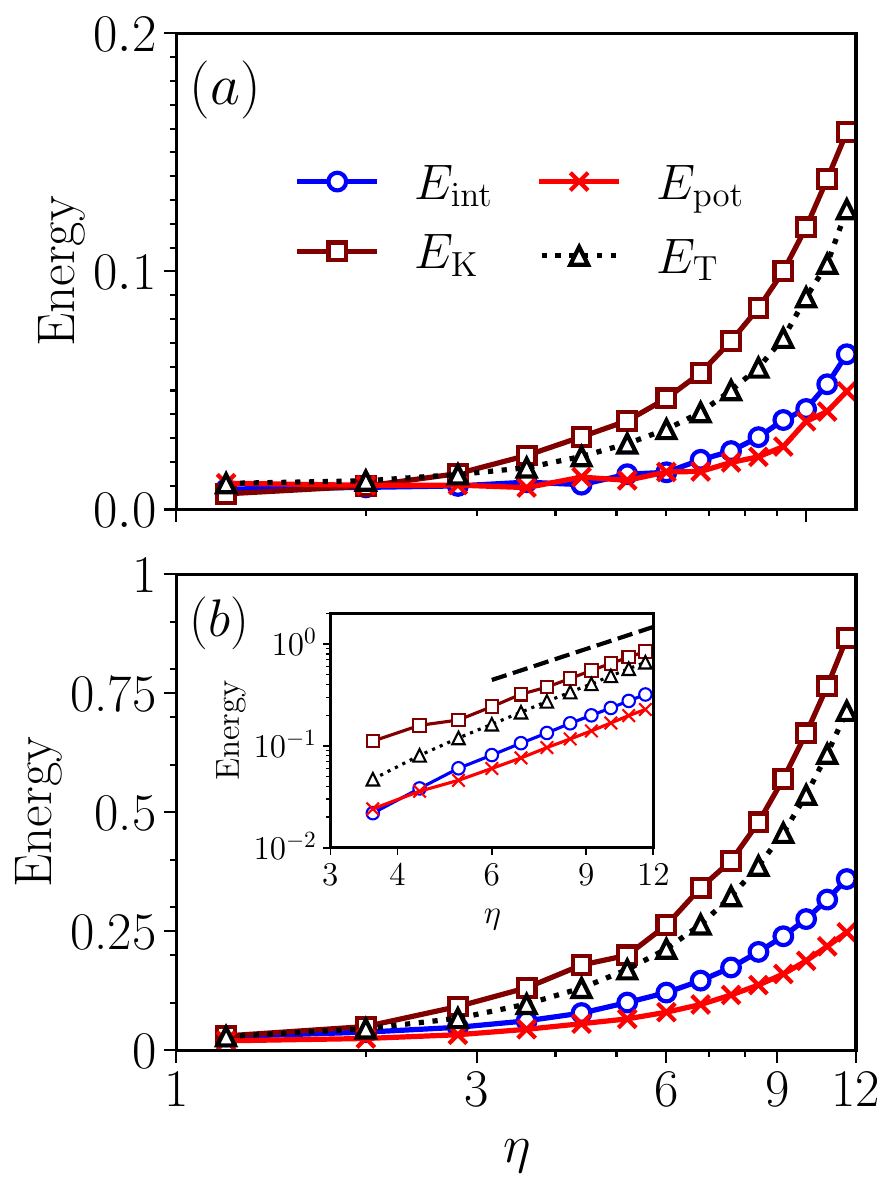}
\caption{Different energies $E_{\rm int}$, $E_{\rm pot}$, $E_{\rm K}$, and $E_{\rm T}$ as a function of $\eta$ for lattice constants: (a) $a_{\rm imp}=2.2$, and (b) $a_{\rm imp}=0.8$. Each energy is scaled as $\rm Energy=\rm Energy/\rm Energy^{0}-1$, where $\rm Energy^{0}$ is the energy of the square vortex lattice without any impurities.} 

\label{fig:energy-static}
\end{figure}
\begin{figure}[!ht]
\centering
\includegraphics[width=\linewidth]{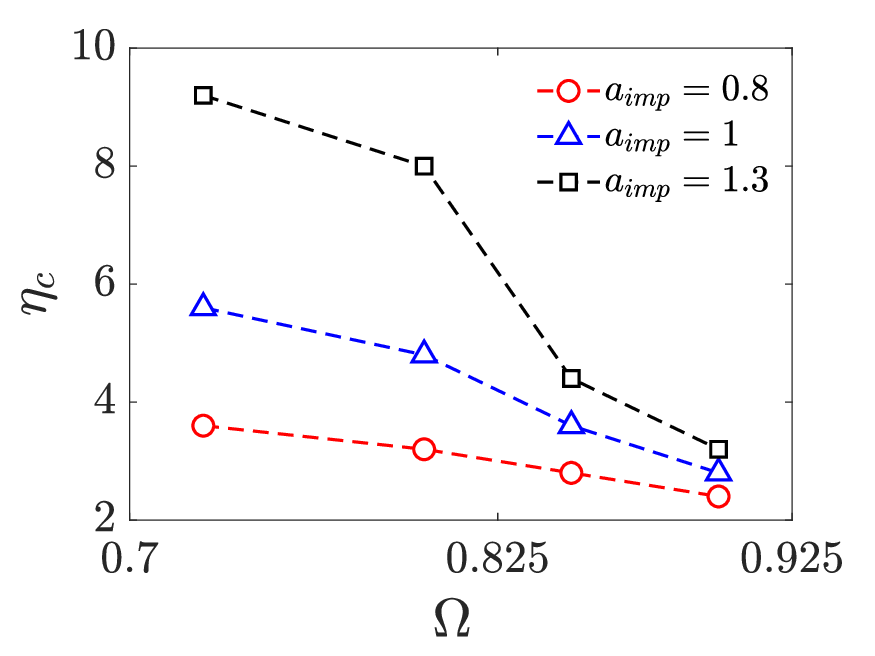}
\caption{Variation of critical impurity strength ($\eta_c$) as a function of $\Omega$ for different impurity lattice constants ($a_{\rm imp}$). The results show that the $\eta_c$ decreases as the $\Omega$ increases.}
\label{fig:etac-php-static}
\end{figure}
Figure~\ref{fig:ang-strct}(b) depicts the angular momentum ($\left\langle L_z \right \rangle$) of the vortex lattice structure as a function of $\eta$ (in a semilog scale). This provides complementary information to the structure factor analysis. We find that $ \left\langle L_z \right\rangle $ remains near zero for $a_{\rm imp}=2.2$, which indicates that no new vortices are created in the condensate at this impurity strength to distort the lattice structure. Conversely, for $a_{\rm imp}=1.3$, $ \left\langle L_z \right\rangle $ exhibits an increasing trend beyond a certain threshold value of $\eta$, suggesting the nucleation of additional vortices. Moreover, as we decrease $a_{\rm imp}$ further, $ \left\langle L_z \right\rangle $ shows a pronounced increasing behaviour with higher magnitudes. The inset of figure \ref{fig:ang-strct}(b) describes the variation of $ \left\langle L_z \right\rangle $ with $\eta$ in the log-log scale which reveals a power law nature of the angular momentum with $\eta$ for larger values of $\eta$ with a critical exponent~$\sim 1.73$. The power law behaviour shows the completely disordered nature of the vortex arrangement. The presence of additional nucleated vortices introduces an incommensurate effect on the vortex lattice and interacts with the pinned vortices. This interaction leads to the deformation of the vortex lattice structure.
\begin{figure}[!ht]
\centering
\includegraphics[width=0.95\linewidth]{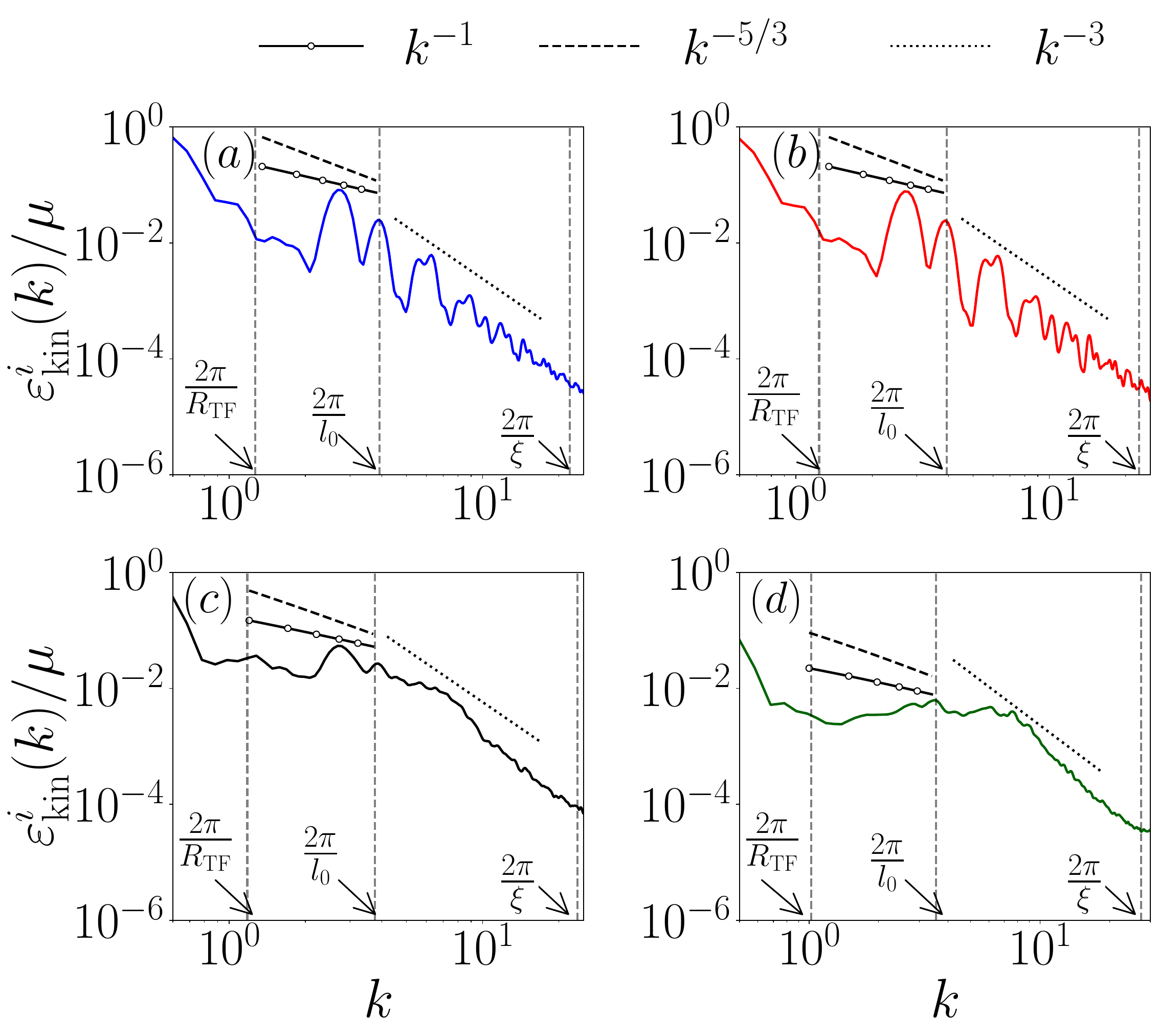}
\caption{Incompressible kinetic energy spectra for different $\eta$ and $a_{\rm imp}$:~(a) $a_{\rm imp}=2.2$, and $\eta=3.6$, (b) $a_{\rm imp}=2.2$, and $\eta=12$, (c) $a_{\rm imp}=0.8$, and $\eta=3.6$, (d) $a_{\rm imp}=0.8$, and $\eta=12$. It exhibits $k^{-1}$ and $k^{-3}$ power laws in the IR and UV regions, respectively.}
\label{fig:incomp_spectra}
\end{figure}

To gain a more comprehensive insight into the emergence of disorder related to vortex lattice melting, we calculate the information entropy of the system. Figure~\ref{fig:entropy-static} illustrates the total information entropy $S_{\rm en}$ of the condensate as a function of the parameter $\eta$, considering different values of $a_{\rm imp}$. For larger values of $a_{\rm imp}$, such as $2.2$, the entropy remains constant regardless of $\eta$. However, as we decrease $a_{\rm imp}$, the entropy shows a noticeable increase with $\eta$. In the logarithmic scale, this increase follows a power-law behaviour with an exponent of $1.73$ in the melted region.


To elucidate the cause of this structural transformation, we calculate the relevant physical contributions by decomposing the total energy into its various components, as shown in equation (\ref{Eq_Energy}). In figure \ref{fig:energy-static}, we show the variation of kinetic~($E_{K}$), potential~($E_{\rm pot}$), interaction~($E_{\rm int}$), and total energy~($E_{T}$) with respect to $\eta$ for different $a_{\rm imp}$. Figure~\ref{fig:energy-static}(a) illustrates that all the energy components are nearly the same order for $a_{\rm imp} = 2.2$. This observation is evident because there is no alteration in the vortex-lattice structure at this specific lattice parameter. However, as we decrease the lattice parameter, the energy components exhibit an increasing trend above a critical value of $\eta$~[see figure \ref{fig:energy-static}(b)]. In the melted vortex lattice region, kinetic energy ($E_{K}$) dominates the other energy components. In the ordered lattice configuration, vortices are arranged in a regular pattern, leading to lower kinetic energy because the motion of vortices is restricted. However, as the vortex lattice melts, the vortices move freely and are no longer confined to fixed positions. Increased disorder facilitates free movement of the vortices, which finally leads to the large kinetic energy. As the vortices are no longer in the pinned state, they interact with each other, and this interaction leads the interaction energy~($E_{\rm int}$) to be dominant over potential energy~(see figure \ref{fig:energy-static}(b)). Additionally, as the value of $a_{\rm imp}$ decreases, the magnitude of the energy components increases. The inset of figure \ref{fig:energy-static}(b) depicts the variation of energy components in the log-log scale and reveals a power law dependence with critical exponent $1.73$ for higher values of $\eta$.

\bgroup
	\def\arraystretch{1.5}
		
		\begin{table}[!hb]
        \centering
		\begin{tabular}{|c||c||c||c||c| } 
		\hline
		\Large $a_{\rm imp}$ &  \Large $\eta$ & \Large $R_{\rm Tf}$ & \Large $\xi$   & \Large $l_0$ \\
		\hline
		\multirow{2}{4em}{\centering $2.2$} & $3.6$ & $4.961$ & $1.605$ & $0.285$ \\ 
		&1$2$  & $5.083$ & $1.594$ & $0.277$   \\ 
		\hline
		
		\multirow{2}{4em}{\centering $0.8$} & $3.6$ & $5.323$ & $1.618$ & $0.265$  \\ 
		& $12$ & $6.19$1 & $1.780$ & $0.228$  \\ 
	 \hline
		\end{tabular}
		\caption{Estimation of different characteristic length scales $R_{\rm TF}$, $l_0$, and $\xi$ for different $a_{\rm imp}$ and $\eta$ at $\Omega=0.8$. All the reported values are averaged over five different realizations of random impurity potential.}
		\label{table:1}
     	\end{table}
\begin{figure}[!ht]
\includegraphics[width=\linewidth]{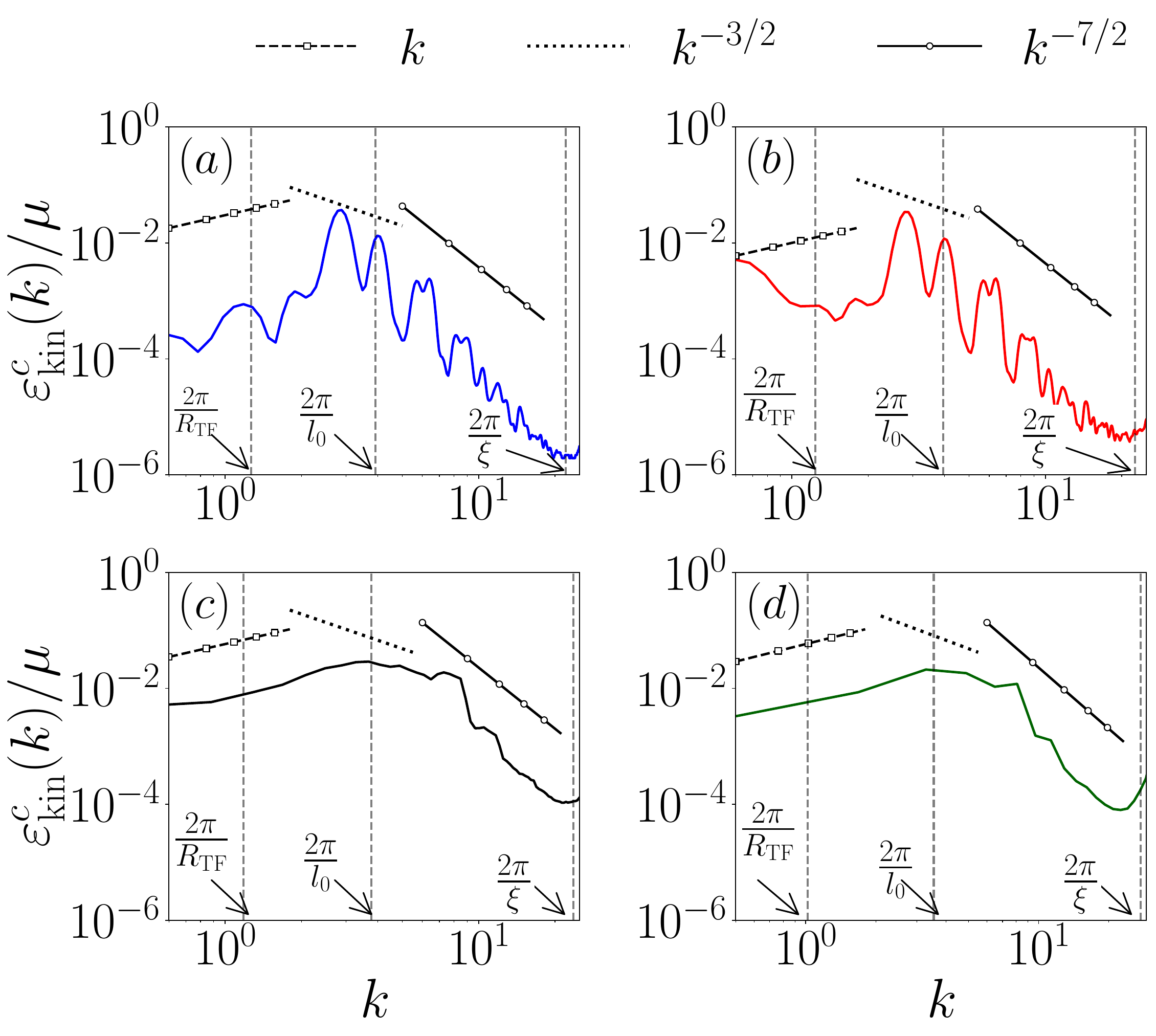}
\caption{Compressible kinetic energy spectra for different $\eta$ and $a_{\rm imp}$:~(a) $a_{\rm imp}=2.2$, and $\eta=3.6$; (b) $a_{\rm imp}=2.2$, and $\eta=12$; (c) $a_{\rm imp}=0.8$, and $\eta=3.6$; (d) $a_{\rm imp}=0.8$, and $\eta=12$. It shows $k$ power law for small value of $k$.  In the UV region, it follows $k^{-3/2}$, and $k^{-7/2}$ power law.}
\label{fig:comp_spectra}
\end{figure}

Next, in Fig.~\ref{fig:etac-php-static}, we present a comparative analysis about the dependence of $\eta_c$ on the rotational frequency $\Omega$ for impurity lattice constants $a_{\rm imp} = 1.3$ (black squares), $1.0$ (blue triangles), and $0.8$ (red circles). We presented the details of this analysis in \ref{appendix}. Our observations indicate that as the rotational frequency increases, the critical value of the strength of the impurity potential above which the vortex lattice melts decreases. The above suggests that higher rotational frequency leads to an earlier onset of vortex lattice melting. Additionally, we notice that the critical value $\eta_c$ increases with the increment of $a_{\rm imp}$ for a lower value of $\Omega$. Although $\eta_c$ appears to be different for a low value of $\Omega$, for higher $\Omega$, $\eta_c$ value converges towards the same order regardless $a_{\rm imp}$.   


To characterize the melted vortex lattice state more comprehensively, we examine the kinetic energy spectra of the condensate. We begin by focusing on the incompressible kinetic energy spectrum ($\varepsilon_{\rm kin}^{i}$) shown in figure \ref{fig:incomp_spectra}, followed by the compressible energy spectra ($\varepsilon_{\rm kin}^{c}$) in figure \ref{fig:comp_spectra}. The scaling laws of these spectra provide a more profound understanding and quantification of the turbulent regions. In our spectral analysis, we identified three distinct length scales in $k$ space, that pertain to the condensate. These length scales are characterized as $k_{R_{\rm TF}}=\frac{2\pi}{R_{\rm TF}}$, $k_{l_0}=\frac{2\pi}{l_0}$, and $k_{\xi}=\frac{\pi}{\xi}$, where $R_{\rm TF}=\sqrt{2\mu}$ represents the Thomas Fermi radius, $l_0=\frac{1}{n_v}$ denotes the inter vortex separation, and $n_v$ stands for the number of vortices per unit area. In our study we decompose the kinetic energy spectrum over wave number $k$ into two different regions: the ultraviolet regime~($k \gg \xi^{-1}$) and the infrared regime~($k \ll \xi^{-1}$). 
In Table~\ref{table:1}, we provide the estimated values of $R_{\rm TF}$, $l_0$, and $\xi$ for different impurity lattice constant $a_{\rm imp}$ and the strength of the impurity potential $\eta$.


Figure~\ref{fig:incomp_spectra} represents the incompressible energy spectra for different impurity lattice constants and the strength of the impurity potential. For $a_{\rm imp} = 2.2$, the spectra show multiple peaks in the range $k_{R_{\rm TF}} < k < k_{l_0}$ for $\eta =3.6$ and $12.0$, indicating the presence of the organized vortex lattice in the condensate. Therefore, no noticeable power law behaviour has been observed in that region [see figure \ref{fig:incomp_spectra}(a) and (b)]. However, the spectrum adheres to a $k^{-3}$ power law in the range $k_{l_0} < k < k_{\xi}$, which is associated with the presence of the vortex core. When the impurity lattice constant is further reduced to $a_{\rm imp}=0.8$, the peaks disappear, and the spectra exhibit power law behaviour with exponents of $k^{-1}$ in the range $k_{R_{\rm TF}} < k < k_{l_0}$ and $k^{-3}$ for larger length scales $k_{l_0} < k < k_{\xi}$ [see figure \ref{fig:incomp_spectra}(c) and (d)]. The scaling behaviour at length scales larger than the inter-vortex distance suggests the presence of turbulence arising from the random motion of the vortices. The presence of $k^{-1}$ scaling of the energy spectrum indicates the Vinen-like cascade of the energy in the scale larger than the intervortex separation. Similar kinds of scalings have been reported for the turbulent two-dimensional condensate trapped in the harmonic trap~\cite{Sivakumar2024}.

Next, we analyze the behaviour of the compressible part of the kinetic energy spectrum in the vortex lattice melted state. In figure~\ref{fig:comp_spectra}, we show the compressible energy spectra $\varepsilon_{\rm kin}^{c}$ for the same set of parameters as shown in figure~\ref{fig:incomp_spectra}. We observe that compressible energy spectra for all impurity lattice constants seem to follow a $k$ scaling for small wave numbers, and this behaviour is related to the thermal equilibrium of the energy. For $a_{\rm imp}=2.2$ [see figure \ref{fig:comp_spectra}(a) and (b)], no distinct power law behaviour has been observed, and the presence of multiple peaks indicates the existence of the vortex lattice structure within the condensate, as discussed earlier in the case of incompressible energy spectra. The spectrum exhibit $k^{-7/2}$ power law in the region $k_{l_0} < k < k_{\xi}$ which corresponds to the turbulence behavior. For high impurity strength, the spectrum develops a $k^{-3/2}$ power law for a small region around $k_{l_0}$~(see figure \ref{fig:comp_spectra}(d)), indicating the weak wave turbulence in melted vortex lattice.

So far, our investigation has been mainly focused on the effect of random impurity potential on the vortex lattice structure, revealing that the lattice constant and amplitude of the impurities significantly affect the structure of the vortex lattice. Further, we show the presence of Vinen-like scaling in the inertial range of the incompressible energy spectrum that suggests the presence of turbulence nature of the condensate in the vortex lattice melted state.  

In the next part of our study, we consider the impurities as a rotating Gaussian obstacle and analyze its effect on the vortex lattice structure and the turbulence nature of the condensate in the melted vortex lattice state.  

\begin{figure}[!ht]
\centering
\includegraphics[width=0.95\linewidth]{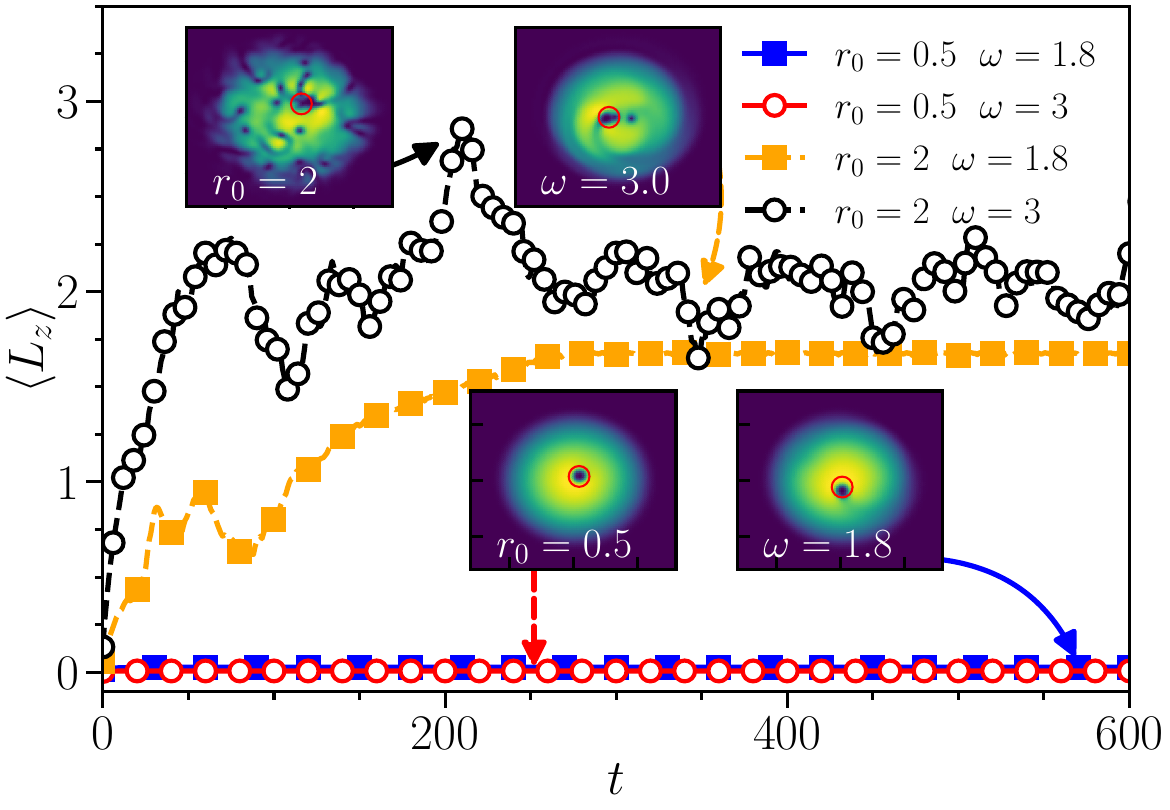}
\caption{Temporal evolution of the $\langle L_z \rangle$ for $r_0= 0.5$ and $2$ at   $\omega$(=1.8, and 2). The inset shows the pseudo-color presentation of the condensate density in real space at $t = 400$. The other parameters are $\Omega = 0.0, ~g = 1000$, and $V_0 = 5.0$. The red circle in the density plot (inset) indicates the obstacle's position. }  
\label{fig:Lz-r0}
\end{figure}
\begin{figure*}[!ht]
\centering
\includegraphics[width=\linewidth]{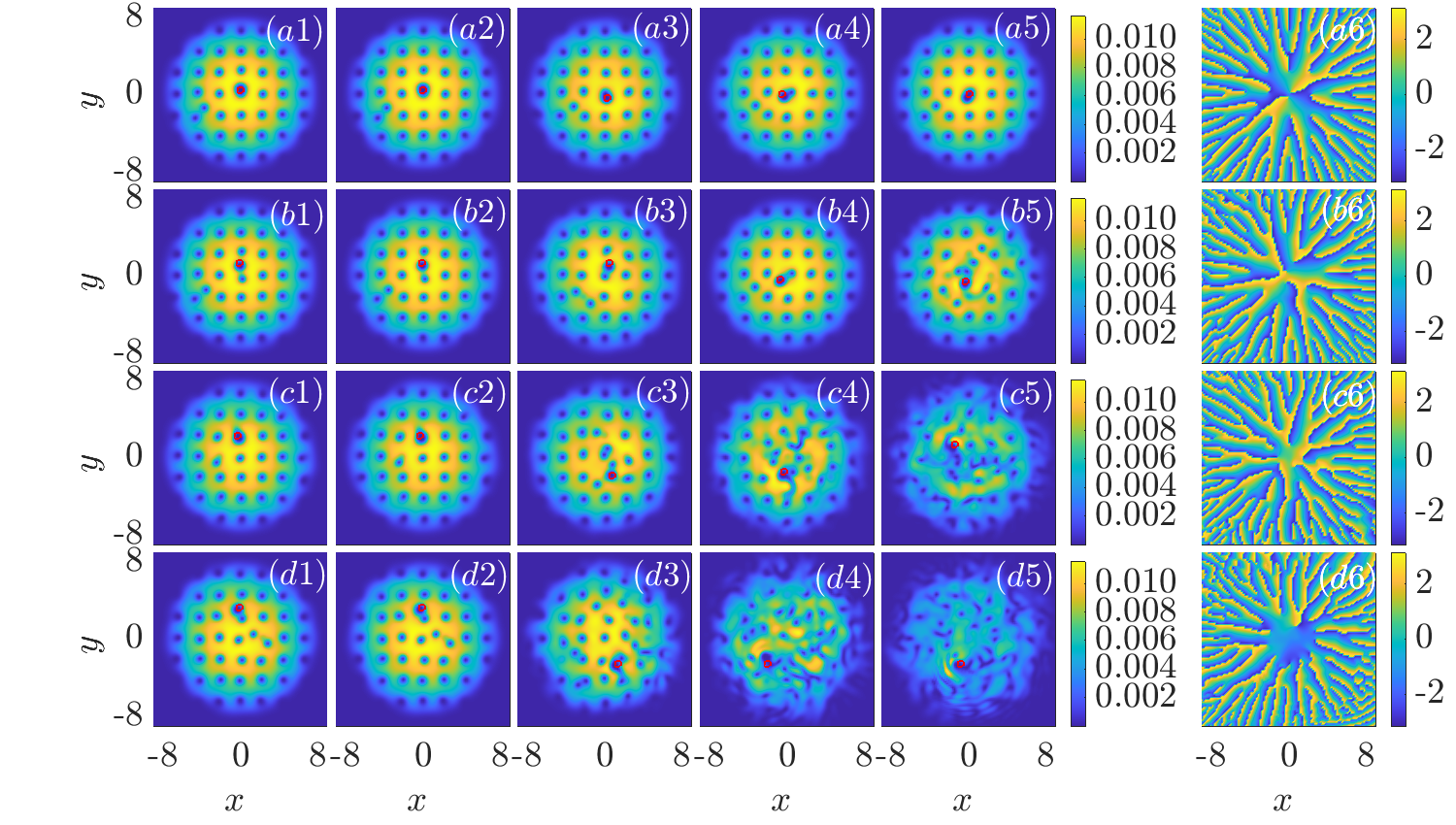}
\caption{Pseudocolour representation of condensate density in real space for different $r_0$ and $\omega$ values of the obstacle. (a1)-(a5): For fixed $r_0=0.5$ and $\omega $ varies ($0, 0.2, 0.6, 1.2$, $2.0$) ; (b1)-(b5): at $r_0=1$ and $\omega = (0, 0.2, 0.6, 1.2$, $2.0 )$ ; (c1)-(c5) at $r_0=1.5$ and $\omega = (0, 0.2, 0.6, 1.2$, $2.0)$; and (d1)-(d5) $r_0=2$ and $\omega =(0, 0.2, 0.6, 1.2, 2)$, respectively. Other parameters are $a=2.2$, $\Omega=0.8$, $V_0=5$ and $g = 1000$. The red circle indicates the obstacle position. The last column (a6-d6) represents the phase of the condensate corresponding to the vortex melting state represented in (a5-d5), respectively.} 
\label{fig_real_obs} 
\end{figure*}

\subsection{Vortex lattice melting in presence of oscillating Gaussian obstacle}
\label{sec:Oscillating}
In this section, we investigate the effect of the oscillating Gaussian obstacle on the structure of the vortex lattice for rotating BECs trapped under a square optical lattice. To probe a systematic way of generation of the vortices in the system we first present the dynamics of the vortices generation in the condensate in the presence of the oscillating Gaussian potential in the absence of rotation and without optical lattice potential. Subsequently, we extend the analysis to investigate the impact of the rotating obstacle on the vortex lattice structure for rotating BECs trapped under a harmonic as well as square lattice potential. 

As the Gaussian obstacle is stirred through the condensate, vortices get generated to minimize the energy of the system~\cite{Reeves:2012, Mithun:2021, Subrata:2022}. There are several ways to control the generation of vortices in an obstacle, either by increasing the distance of the obstacle $r_0$ from the centre of the condensate or by increasing the oscillation frequency $\omega$ of the obstacle. Here, we aim to investigate the resultant dynamical effect of the obstacle on the vortex lattice state. To make the analysis self-contained, in the following, we concisely present the dynamics of non-rotating condensate in the presence of an oscillating obstacle for various $r_0$ and $\omega$. 

Following earlier work, we consider an impenetrable obstacle with $V_0/\mu~>1$~\cite{Mithun:2021} with obstacle depth $V_0=80$ and the chemical potential $\mu=12.26$. The width of the obstacle has been fixed to $d = 0.3$.

\begin{figure*}[!ht]
\centering
\includegraphics[width=0.85\linewidth]{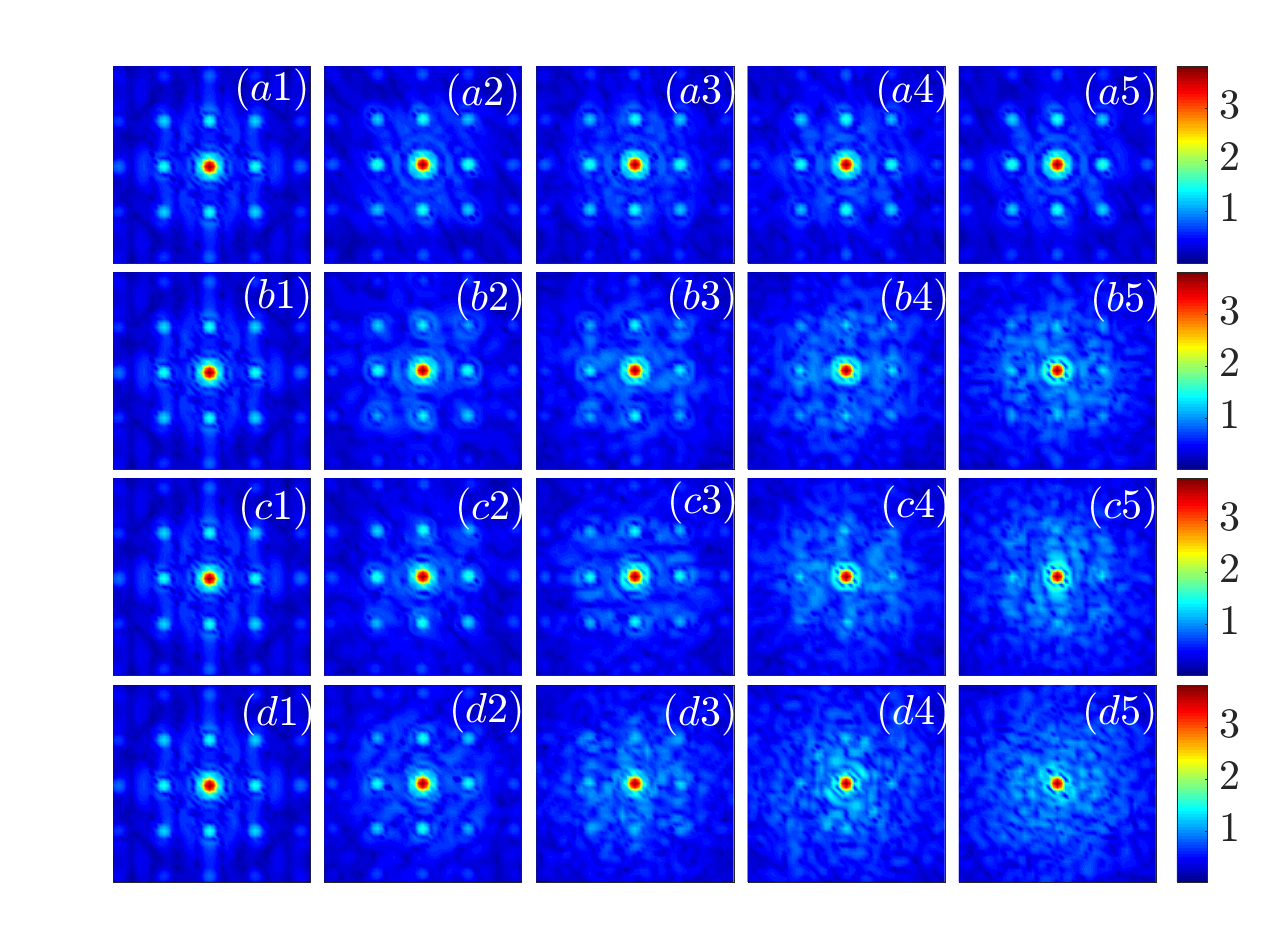}
\caption{Pseudocolour representation of density in momentum space corresponding to the real space densities in figure \ref{fig_real_obs}. (a1-a5): For $r_0=0.5$ and $\omega = (0.2, 0.6, 1.2$, $2$); (b1-b5): For fixed $r_0= 1.0$ and $\omega = (0.2, 0.6, 1.2$, $2.0$); (c1-c5) For $r_0 = 1.5$ and $\omega = (0.2, 0.6, 1.2$, $2$) ; and (d1-d5) For $r_0=2$ and $\omega = (0.2, 0.6, 1.2$, $2.0$). The other parameters are the same as in figure \ref{fig_real_obs}.}
\label{fig_k_obs}
\end{figure*}

In figure \ref{fig:Lz-r0}, we show the temporal evolution of the angular momentum $\left\langle L_z \right\rangle=\int \psi^{\dagger} \left( y \partial_x - x \partial_y \right) \psi dx dy$, for different distance of the obstacle from the centre as well as for various frequencies keeping the $r_0$ fixed. We consider $r_0 = 0.5$ (depicted by red open circles) and $r_0 = 2.0$ (black open circles) at $\omega = 1.8$, as well as $\omega = 1.8$ (depicted by blue filled squares) and $\omega = 3.0$ (orange filled squares) at $r_0 = 1.0$. For the case of $r_0 = 0.5$, the angular momentum $\left\langle L_z \right\rangle$ remains zero indicating the absence of the vortices. However, as the value of $r_0$ is increased to 2.0, we observe an increase in the angular momentum for the initial time, which subsequently fluctuates within the range $1.5 \lesssim \left\langle L_z \right\rangle \lesssim 2.5$ over after a long time ($t\gtrsim 300$). The presence of a non-zero angular momentum indicates the generation of vortices within the condensate. 

Similarly, for $\omega = 1.8$, the angular momentum remains at zero, indicating the absence of vortices. In contrast, for $\omega = 3.0$, the angular momentum follows a similar trend for $r_0 = 2.0$. Interestingly, after $t \geq 230$, the angular momentum reaches a value of approximately $2.0$ in the case of single vortex generation ($\omega = 3.0$) or the case of multiple vortex generation at $r_0 = 2.0$. This behaviour is visually evident from the inset density profile. 


So far, we have observed that vortex generation by dynamic impurities modelled as Gaussian obstacles can be influenced by two factors: the distance from the condensate center ($r_0$) and the oscillation frequency ($\omega$) of the obstacle. Next, we consider the rotating BECs under the square optical lattice and harmonic trap and analyze the impact of the dynamic Gaussian impurities on the structure of the vortex lattice by varying its frequency and its position from the centre of the condensate in the similar line as studied with the static impurities case in the section~\ref{sec:vortexMelt}. In figure \ref{fig_real_obs}, we show the pseudo colour representation of the density while varying both $r_0$ and $\omega$ of the dynamical Gaussian obstacle. We find that at $r_0=0.5$, the vortex lattice structure remains pinned even for higher oscillation frequencies [see figure \ref{fig_real_obs}(a1)-(a5)]. However, we notice a small distortion in the vortex lattice structure at $\omega=3$ upon increasing $r_0$ from $0.5$ to $1$ [see figure \ref{fig_real_obs}(b5)]. The vortex lattice structure starts getting deformed quite significantly around $\omega=1.2$ upon considering $r_0 = 1.5$ as shown in figure \ref{fig_real_obs}(c4). From figure \ref{fig_real_obs}(d2), we can see that an increase in $r_0$ from $1.5$ to $2$ results in deformation in the lattice structure even at lower oscillation frequency $\omega=0.6$. These characteristics are more pronounced in the density distribution in momentum space [see figures \ref{fig_k_obs}(a1)-(a5)]. We find that the periodic peaks are arranged in a square pattern, signifying the pinned vortex lattice of the condensate density in real space. In the presence of a sufficiently high oscillation frequency and high $r_0$, the translational symmetry and long-range order of the lattice structure are completely destroyed. As a result, no periodic peaks are visible, while the vortex lattice structure remains in the square lattice. Thus, it characterized the melted state of the vortex lattice structure, which appears in figures \ref{fig_k_obs}(d1)-(d5). We show the the structural transformation of the vortex lattice from an ordered to a disordered state based upon the values of $\omega$ and $r_0$ in more systematic way in the figures \ref{fig_time_strct_obs} and \ref{fig_avg_strct_obs}. 
\begin{figure}[!ht]
\includegraphics[width=\linewidth]{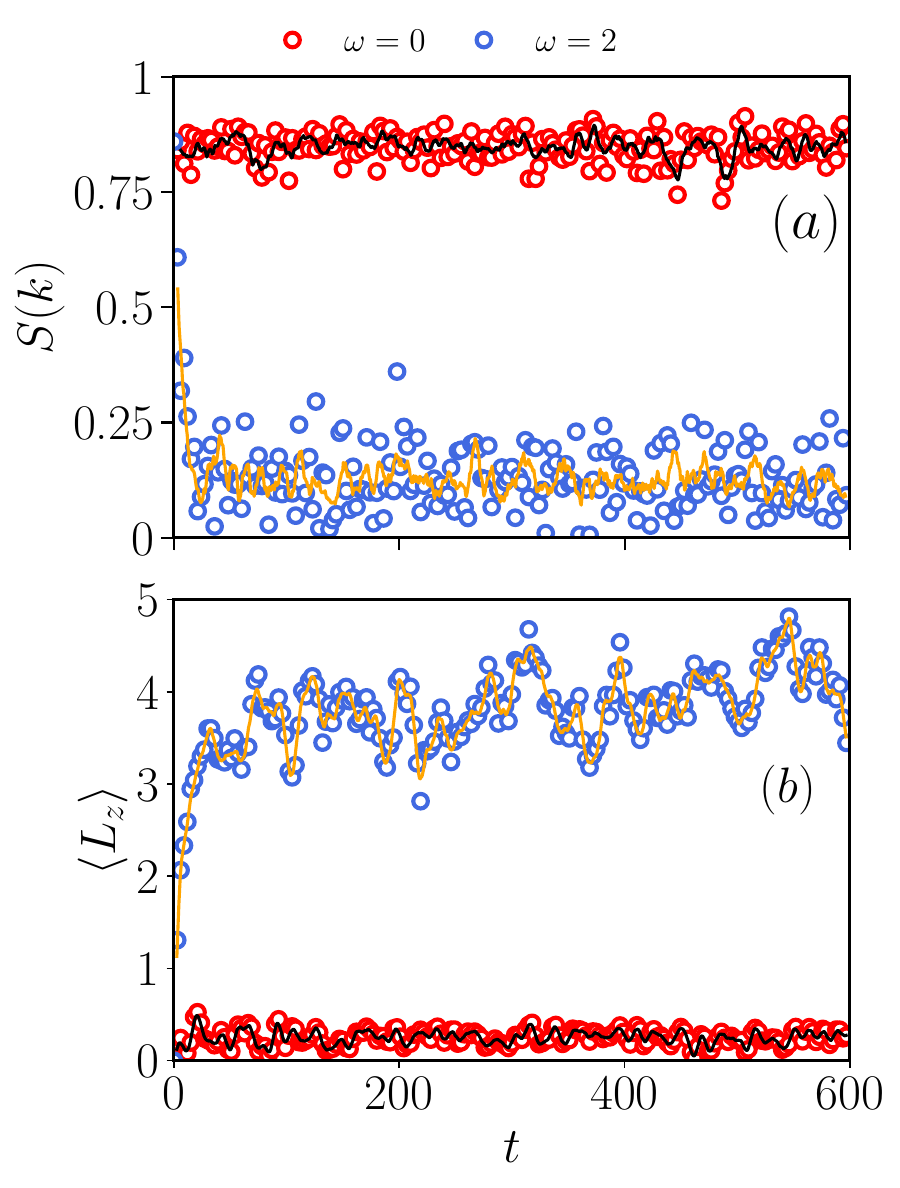}
\caption{Temporal evolution of (a) the structure factor ($S(k)$), and (b) the angular momentum ($\left\langle L_z \right\rangle$) of the condensate for different oscillation frequencies ($\omega = 0.0, 2.0$) for the obstacle located at $r_0 = 2.0$. The black and orange solid lines represent the moving average at $\omega = 0.0$ and $\omega = 2.0$, respectively. } 
\label{fig_time_strct_obs}
\end{figure}
\begin{figure}[!ht]
\centering
\includegraphics[width=\linewidth]{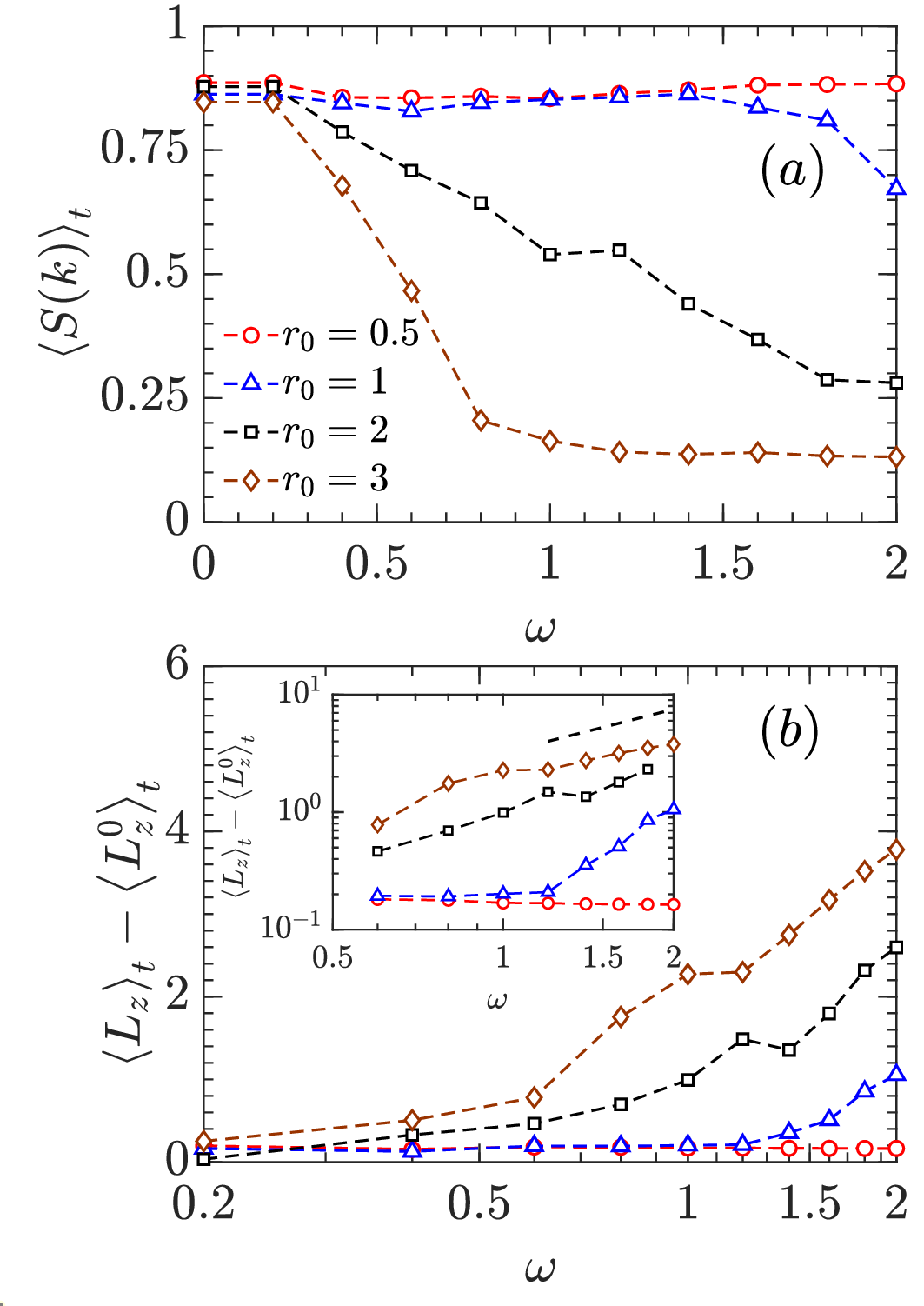}
\caption{(a) Time average of structure factor $\left\langle S(k) \right\rangle_{t}$ and (b) angular momentum $\left\langle L_{z} \right\rangle_{t} - \left\langle L^0_z \right\rangle_{t}$ as a function $\omega$ for different $r_0$. The inset in (b) shows the variation $\left\langle L_{z} \right\rangle_{t} -\left\langle L^0_z\right\rangle_{t}$ in the log scale. $\left\langle L^0_z \right\rangle_{t}$ is the angular momentum of the initially pinned vortex lattice at $r_0=0$, and $\omega=0$. The black dashed line is drawn to show the power law nature with $\omega$ in the melted region.}
\label{fig_avg_strct_obs}
\end{figure}

Similar to the analysis of static impurities case, here also to quantify the structural transformation of the vortex lattice in the presence of the impurities, we compute the structure factor ($S(k)$) and the angular momentum ($\left\langle L_z \right \rangle$) of the condensate, taking into account the square lattice configuration. Figure~\ref{fig_time_strct_obs} illustrates the temporal variations of $S(k)$ and $\left\langle L_z \right \rangle$ for different oscillation frequency $\omega$ at $r_0=2$. We find that when $\omega=0$ (red open circles), $S(k)$ remains relatively constant around $0.8$. However, increasing $\omega$ to $2.0$ induces a deformation in the vortex lattice structure, leading to a significant decrease in $S(k)$ towards 0 (blue open circles). For small values of $\omega$, no interstitial vortices are present in the condensate, thereby causing $\left\langle L_z \right \rangle$ to remain close to 0 (red open circles). Conversely, for larger values of $\omega = 2.0$, the angular momentum $\left\langle L_z \right \rangle$ exhibits an increasing trend over time as depicted in figure~\ref{fig_time_strct_obs}. This feature indicates that extra vortices are being nucleated within the condensate due to the faster oscillation of the obstacle, and these newly generated vortices interact with the vortices pinned near the maxima of the optical lattice, which eventually results in the melting of the vortex lattice structure~\cite{Mithun:2021}.

To have more insight into the vortex lattice melting state in the presence of the dynamical impurities, we analyze the time-averaged structure factor $\left\langle S(k) \right\rangle_{t}$ and angular momentum $\left\langle L_z\right\rangle_{t} - \left\langle L^0_z \right\rangle_{t}$ as a function of $\omega$ for different $r_0$. Here, $\left\langle L_z \right\rangle_{t} $ is rescaled with respect to the angular momentum of the initial state $\left\langle L^0_z \right \rangle_{t}$ at $\omega = 0$. Notably, a significant decrease in $\left\langle S(k) \right\rangle_{t} \lesssim 0.5$ indicates the structural deformation of the vortex lattice, consistent with the observed density profiles. For $r_0 = 0.5$, the $\left\langle S(k) \right\rangle_{t}$ remains approximately 0.85. However, as we increase $r_0$ to $1$, $\left\langle S(k) \right\rangle_{t}$ decreases. Moreover, when $r_0 = 2$, the critical oscillation frequency at which the vortex lattice structure begins to exhibit melting behaviour is $\omega \sim 1.2$, while for $r_0 = 3$, the threshold value is approximately $\omega \sim 0.8$. Thus, an increase in $r_0$ leads to a decrease in the critical $\omega$ beyond which the vortex lattice structure undergoes melting. In figure~\ref{fig_avg_strct_obs}(b), we observe that $\langle L_z \rangle_{t}$ remains close to its initial value for smaller $r_0$ but increases as $r_0$ becomes larger. The increase of the angular momentum value beyond the threshold value of $\omega$ indicates the generation of additional vortices in the condensate, with the number of generated vortices depending on $\omega$ and $r_0$. In the inset of figure~\ref{fig_avg_strct_obs}(b), we show $\left \langle L_z \right \rangle$ in a logarithmic scale, revealing a power-law behaviour with $\omega$ in the melting region, characterized by an exponent of $1.73$ which is same as the case for static impurity.
\begin{figure}[!ht]
\centering
\includegraphics[width=\linewidth]{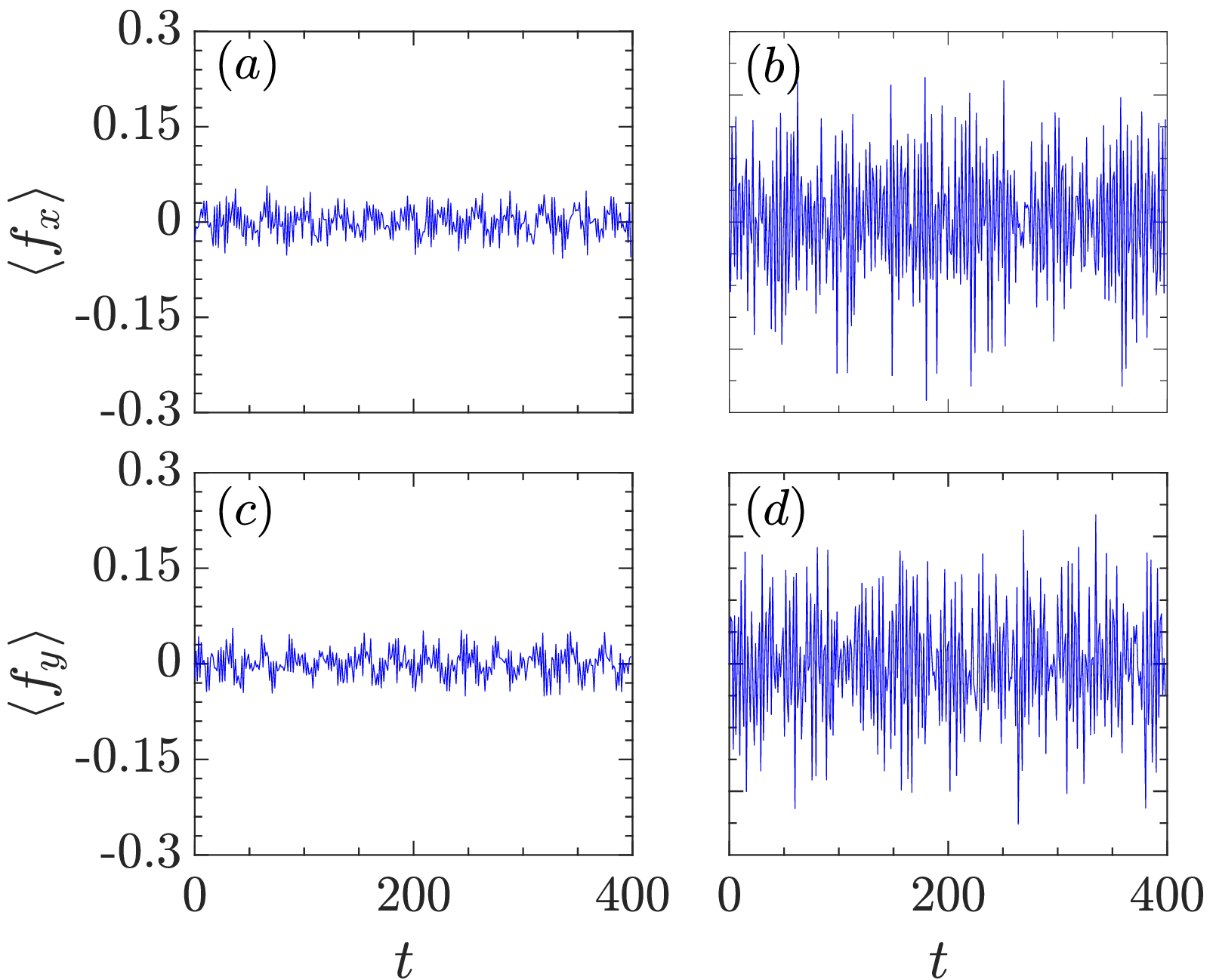}
\caption{Time evolution of $\left\langle f_x \right\rangle$ and $ \left\langle f_y \right\rangle$ with $r_0=2$ for different values of $\omega$. Panels (a) and (b) display the evolution of $\left\langle f_x \right\rangle$, while panels (c) and (d) depict the evolution of $\left\langle f_y \right\rangle$ for $\omega=0$ and $\omega=2$, respectively. When $\omega=0$, both $\left\langle f_x \right\rangle$ and $\left\langle f_y \right\rangle$ oscillate around zero. In contrast for $\omega=2$, they increase periodically.} 
\label{fig_drg}
\end{figure}
\begin{figure}[!ht]
\centering
\includegraphics[width=\linewidth]{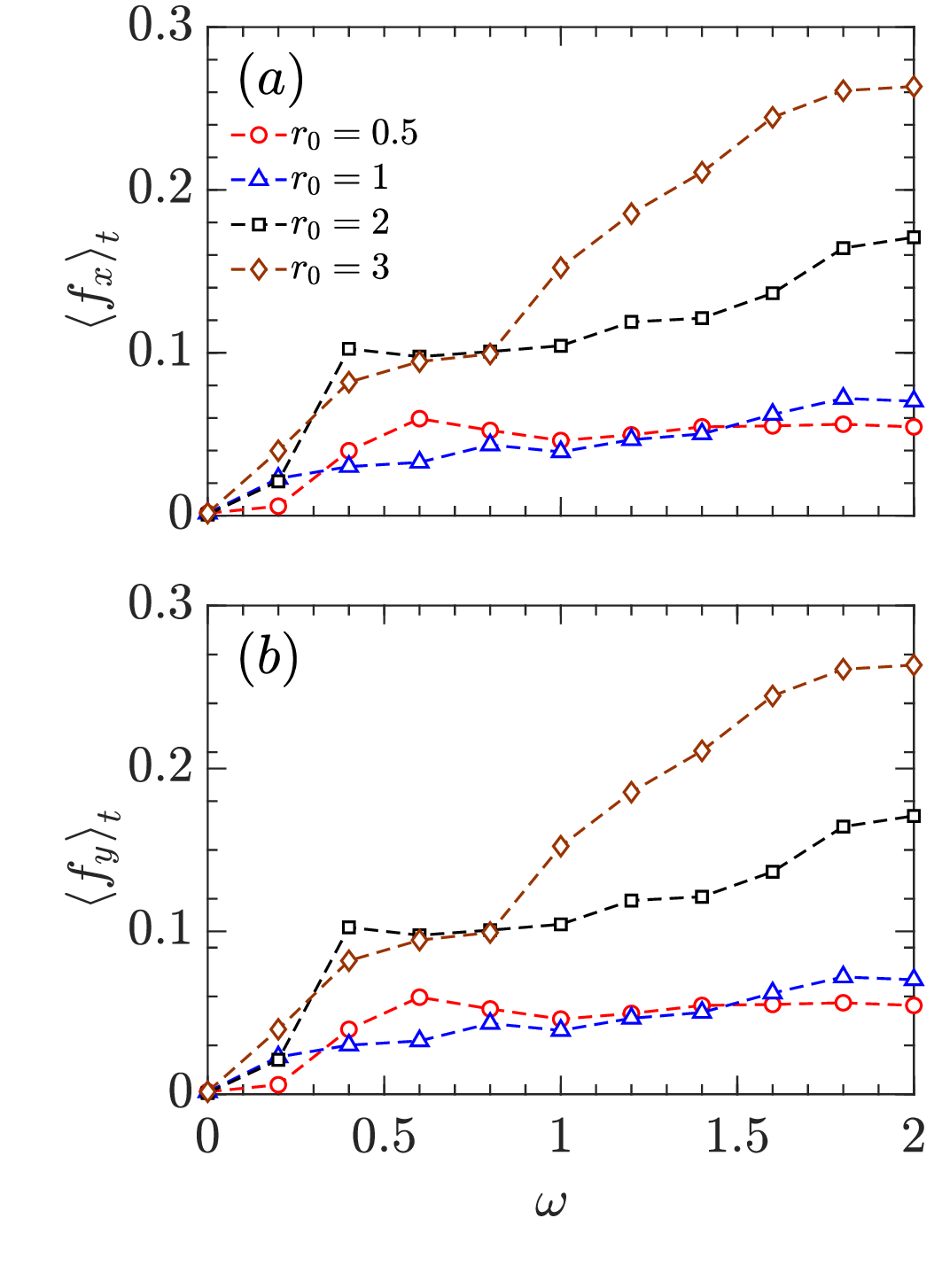}
\caption{Time-average of drag force with oscillation frequency $\omega$ for (a) $\left \langle f_x \right\rangle$ and  (b) $\left \langle f_y \right\rangle$ as a function of $\omega$ for various values of $r_0$. Both $\left \langle f_x \right\rangle$ and $\left \langle f_y \right\rangle$ increase with $\omega$ for larger values of $r_0$.} 
\label{fig_drg_avg}
\end{figure}
To investigate the influence of vortex generation in a condensate, we analyze the normalized drag force acting on an oscillating obstacle which is defined as $\vec{f}=(\left\langle f_x \right\rangle, \left\langle f_y \right\rangle)=i\partial_{t}\int \psi^{\dagger}\nabla \psi dx dy$~\cite{li:2019}. In figure~\ref{fig_drg} we show the x- and y-component of the drag force for $r_0=2$ at $\omega=0$, and $2$. The drag force serves as an indicator of the impact of vortices on the dynamics of an obstacle within a condensate. When vortices are immobilized at their respective positions in the optical lattice, the drag force is relatively small compared to the case where vortices are randomly distributed throughout the condensate. For $\omega=0$ [figure \ref{fig_drg}(a) and figure \ref{fig_drg}(c)], where no additional vortices are nucleated in the condensate, the drag force components $\left\langle f_x \right\rangle$ and $\left\langle f_y \right\rangle$ remain zero over time. However, when additional vortices are generated, $\left\langle f_x \right\rangle$ and $\left\langle f_y \right\rangle$ exhibit oscillatory behavior within the range of $-0.15 \lesssim (\left\langle f_x \right\rangle, \left\langle f_y \right\rangle) \lesssim 0.15$ [figure \ref{fig_drg}(b) and figure \ref{fig_drg}(d)]. Furthermore, we calculate the time average ($\left\langle f_{(x,y)} \right\rangle_{t} = \sum_{i=0}^T \left\langle f_{(x,y)} \right\rangle/T$) of $\left\langle f_x \right\rangle$ and $\left \langle f_y \right\rangle $ to analyze their variation with the oscillation frequency $\omega$ and the position of the obstacle $r_0$, as depicted in figure~\ref{fig_drg_avg}. It is evident that the averaged values $\left\langle f_x \right\rangle_{t}$ and $\left\langle f_y \right\rangle_{t}$ initially increase, but at large $\omega$, the rate of increase becomes significantly lower for low $r_0$. On the other hand, $\left\langle f_x \right\rangle_{t}$ and $\left\langle f_y \right\rangle_{t}$ gradually increase as the vortex lattice structure undergoes a transition from a pinned lattice to a melted lattice structure. Additionally, for higher values of $r_0$, the magnitude of the drag force dominates over lower $r_0$ values for a given $\omega$. In conclusion, the drag force also complements the findings of the structural transformation by showing the dependence of both $r_0$ and $\omega$.

Now, we move on to estimating the extent of vortex disorder using entropy. In figure~\ref{fig_entropy_obs} we plot the time average of total information entropy~($\left\langle S_{\rm en}\right\rangle_{t}$) of the condensate as a function of $\omega$ . The non-zero value of the information entropy indicates the disorderness in the system. It is easy to see that for smaller values of $r_0$ such as $0.5$, the entropy remains nearly constant across various $\omega$. This particular feature suggests that the vortex lattice structure remains relatively ordered, and there is minimal disruption in its arrangement. The entropy maintains a consistent level, suggesting a well-organized state. However, as we increase $r_0$ to larger values, specifically $2$ and $3$, we notice the emergence of a distinctive trend. The entropy depicts a noticeable increase with higher $\omega$ as the larger oscillation frequency disrupts the ordered arrangement of the vortices, leading to a significant rise in the condensate entropy. In the melted region, we observe a power law behaviour in $\left\langle S_{en}\right\rangle_{t}$ with $\omega$ similar to the angular momentum with an exponent of $1.73$ as depicted in the inset of figure. In order to provide a comprehensive illustration of the structural transformation of vortices, we present a phase diagram in the $r_0 - \omega$ plane based on the structure factor analysis. The phase diagram, depicted in figure \ref{fig_phase_obs}, reveals two distinct regions, the square vortex lattice (SVL) and melted vortex lattice (MVL) region. It is evident that the lattice structure remains unaltered for lower values of $r_0$ even for higher $\omega$, and the threshold value of $\omega$ decreases with increasing $r_0$. 

\begin{table}[!b]
\centering
	\begin{tabular}{|c||c||c||c||c| } 
		\hline
		\Large $r_0$ &  \Large $\omega$ & \Large $R_{\rm Tf}$ & \Large $\xi$   & \Large $l_0$ \\
		\hline
		\multirow{2}{4em}{\centering $2$} & $0.6$ & $5.003$ & $1.567$ & $0.282$ \\ 
		& $2$  & $5.145$ & $1.542$ & $0.274$   \\ 
		\hline
		
		\multirow{2}{4em}{\centering $3$} & $0.6$ & $5.093$ & $1.571$ & $0.277$  \\ 
		& $2$ & $5.221$ & $1.515$ & $0.271$  \\ 
		\hline
	\end{tabular}
	\caption{Estimation of different characteristic length scales $R_{\rm TF}$, $l_0$, and $\xi$ for different $r_0$ and $\omega$ at $\Omega=0.8$. All the reported values are averaged over different time intervals, specifically at $t=0, 50, 100, 200, 300$, and $400$.}
		\label{table:2}
\end{table}   	     	

\begin{figure}
\centering
\includegraphics[width=\linewidth]{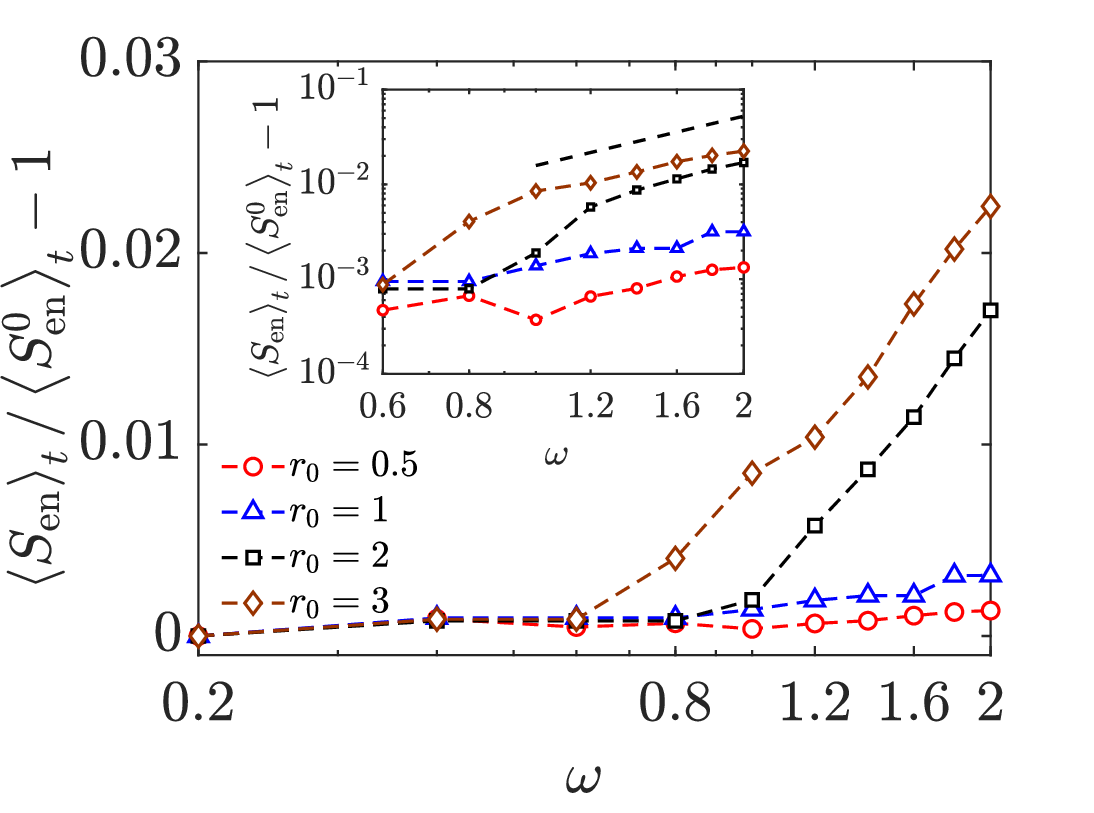}
\caption{Variation of the time average of total entropy~$\left\langle S_{en} \right\rangle_{t}$ with obstacle oscillation frequency ($\omega$) for different values of $r_0$ in semilog scale. The inset figure shows the variation of $\left\langle S_{en} \right\rangle_{t}$ in log scale, which displays a power-law behaviour for larger $\omega$. Here, the entropy rescaled with the entropy of the pinned lattice $\langle S_{en}^{0} \rangle_{t}$ at $V_0=5$.} 
\label{fig_entropy_obs}
\end{figure}

\begin{figure}[!ht]
\centering
\includegraphics[width=\linewidth]{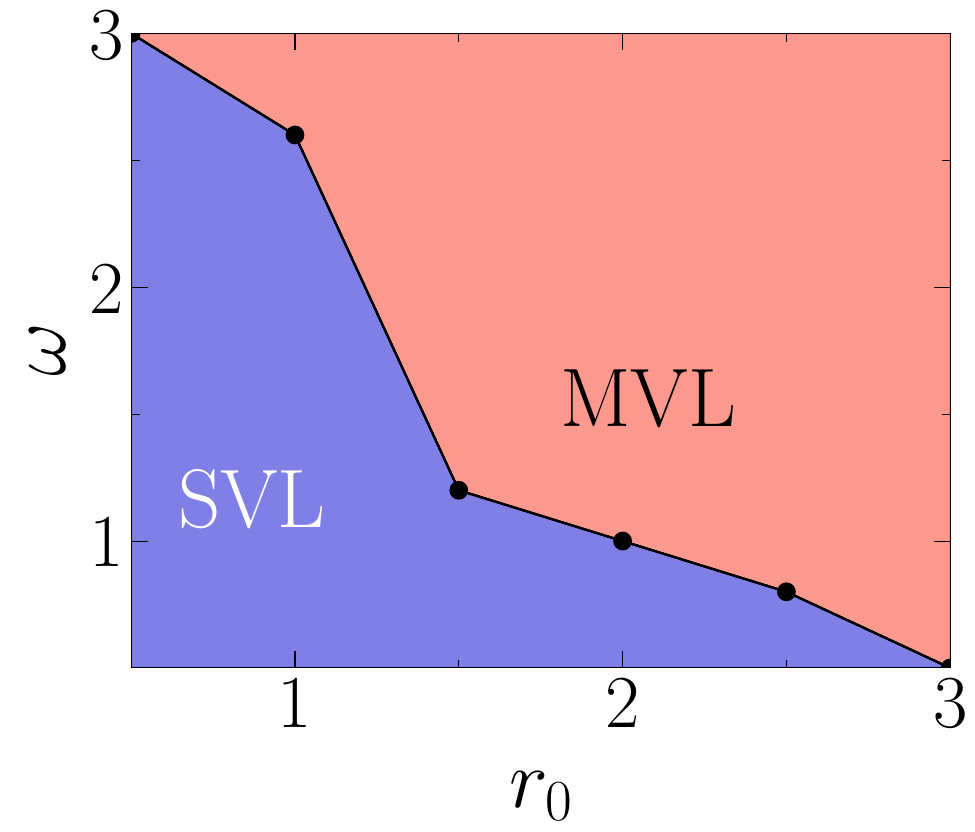}
\caption{Phase diagram of vortex lattice structures in $r_0$-$\omega$ plane at $\Omega=0.8$. The blue-shaded region represents the square vortex lattice (SVL), and the region filled with red colour represents the melted vortex lattice (MVL) region. The black dashed line was obtained using the $\langle S(k) \rangle$ values to differentiate the two regions.} 
\label{fig_phase_obs}
\end{figure}

\begin{figure}[!ht]
\centering
\includegraphics[width=0.95\linewidth]{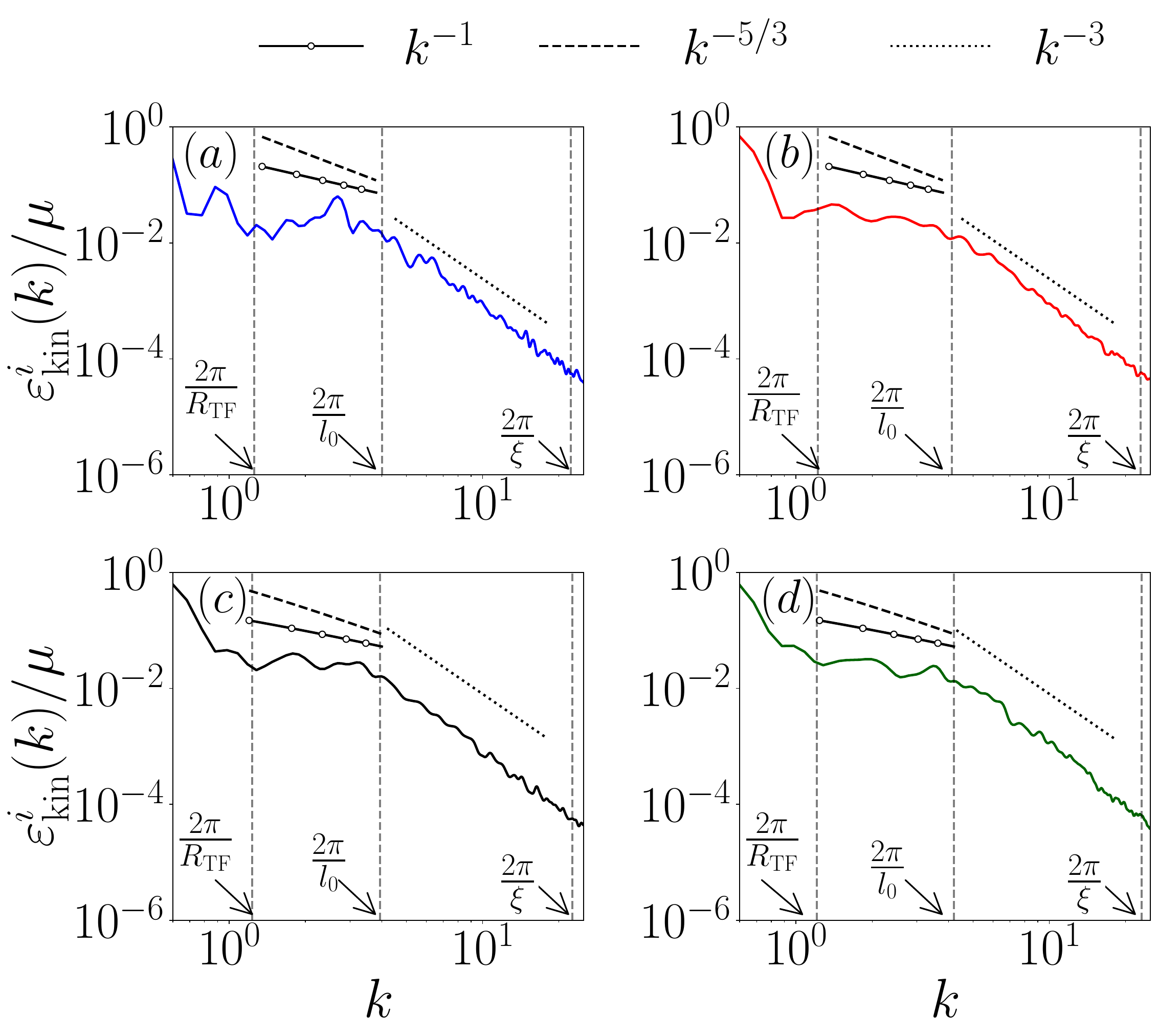}
\caption{Incompressible kinetic energy spectra for different $r_0$, and $\omega$: (a) $r_0=2, \omega=0.6$; (b) $r_0=2, \omega=2$; (c) $r_0=3, \omega=0.6$; (d) $r_0=3, \omega=2$. It follows $k^{-1}$ and $k^{-3}$ power laws in the IR and UV regions, respectively. }
\label{fig_incomp_spectra_obs}
\end{figure}

\begin{figure}[!ht]
\includegraphics[width=\linewidth]{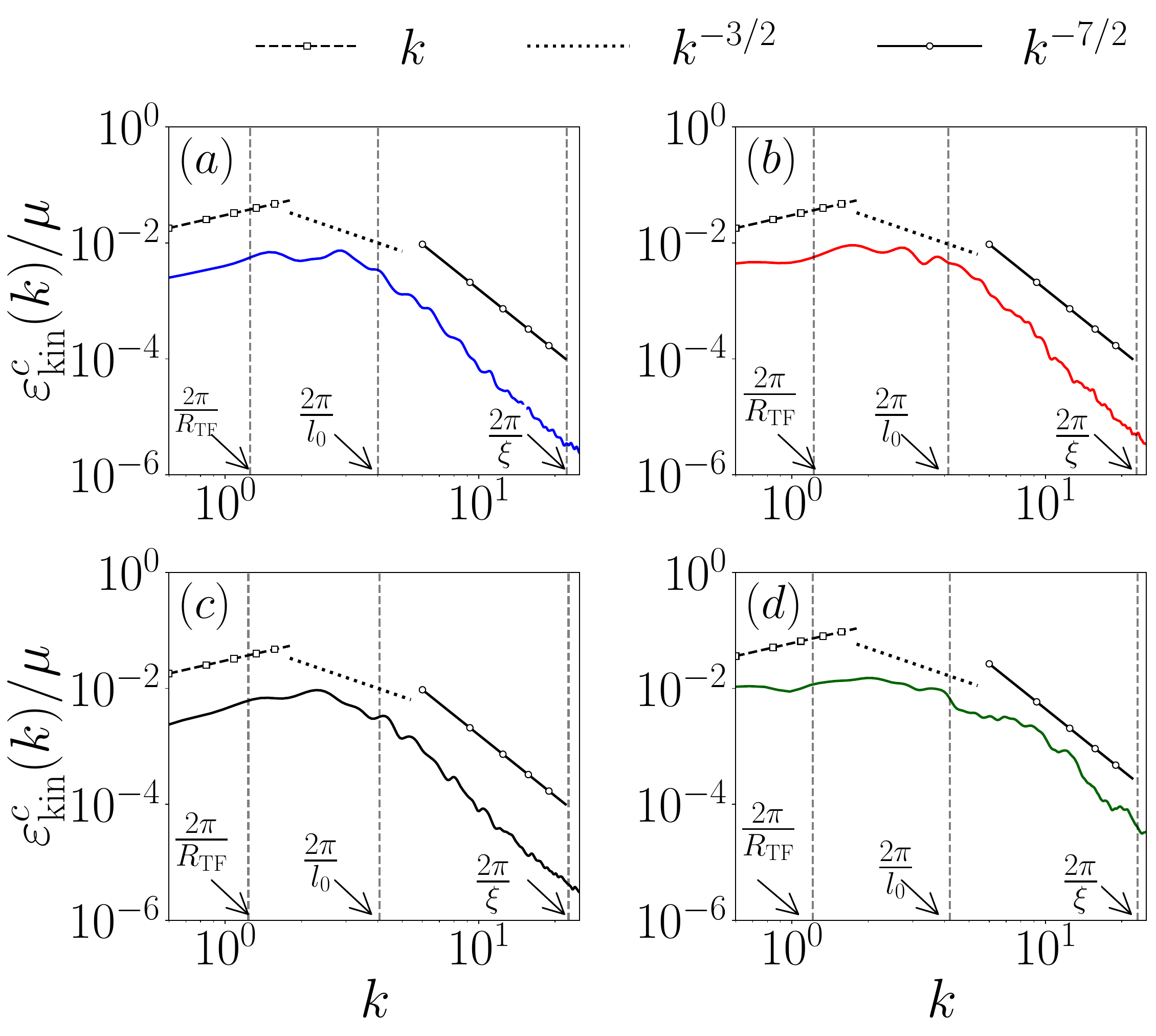}
\caption{Compressible kinetic energy spectra for different $r_0$, and $\omega$: (a) $r_0=2$, and $\omega=0.6$; (b) $r_0=2$, and $\omega=2$; (c) $r_0=3$, and $\omega=0.6$; (d) $r_0=3$, and $\omega=2$. It shows $k$ power law for small value of $k$.  In the UV region, it exhibits $k^{-3/2}$, and $k^{-7/2}$ power law.}
\label{fig_comp_spectra_obs}
\end{figure}

Next, we examine the turbulent behaviour associated with vortex lattice melting by incorporating the incompressible ($\varepsilon_{\rm kin}^{i}$) and the compressible ($\varepsilon_{\rm kin}^{c}$) energy spectrum of the condensate. Similar to the random impurity potential discussed in the previous section, we identify three distinct length scales, namely $k_{R_{\rm TF}}$, $k_{l_0}$, and $k_{\xi}$ in $k$ space. We compute both the compressible and incompressible spectra at different time intervals, specifically at $t=0, 50, 100, 200, 300$, and $400$. We calculate the average energy spectra for both compressible and incompressible components. In figure \ref{fig_incomp_spectra_obs}, we plot $\varepsilon_{\rm kin}^{i}$ for various $r_0$ and $\omega$. Similar to the random impurity potential for all $r_0$, the spectrum displays a power-law behaviour $k^{-3}$ at the length scale greater than $k_{\xi}$, primarily attributed to the vortex core. We find the presence of Vinen-like scaling ($\varepsilon_{\rm kin}^{i}\sim k^{-1}$)  in the region $k_{R_{\rm TF}} < k < k_{l_0}$, characterized by the clustering of vortices without combining into a larger vortex~\cite{Sivakumar2024, Marino:2021}. In figure \ref{fig_comp_spectra_obs}, we show the compressible energy spectra, utilizing the same parameters outlined previously in figure \ref{fig_incomp_spectra_obs}. The observed spectra demonstrate a specific trend characterized by $k$ for small wave numbers. In the range $k_{R_{\rm TF}} < k < k_{l_0}$, the spectra display a power law scaling of $k^{-3/2}$ a typical feature of the weak wave turbulence as reported in~\cite{Sivakumar2024}.  However, for higher wave number range, i.e., $k_{l_0} < k < k_{\xi}$, the spectra show the $k^{-7/2}$ scalings.

\section{\label{conclusion} Conclusions and Perspective} 
In this work, we have investigated the effect of impurities on the structural transformation, leading to the turbulence state of a vortex lattice structure appearing as a ground state for the rotating condensate trapped under the square optical lattice. We have considered two types of impurities, namely, random static impurities and dynamical ones generated using the oscillating Gaussian obstacles. With static impurities have considered the effect of both the lattice constant of the impurity potential as well as its strength. We have found that the square vortex lattice structure gets melted upon increasing the strength of the impurities beyond a threshold strength of the potential. The threshold strength of the potential beyond which the vortex lattice melts decreases upon decreasing the lattice constant of the impurities below the lattice constant of the vortex lattice. We have also established a critical role of the rotational frequency of the condensate in influencing the critical impurity strength required for the vortex lattice melting. To characterize the melted state of the vortex lattice more concretely, we have computed the incompressible and compressible kinetic energy spectrum, which display the presence of turbulence-like features in the condensate power with the presence of Vinen-like scaling of the spectra in the infrared regime.

To show the presence of the universality in the melted state of the vortex lattice structure, we have also investigated the influence of an oscillating Gaussian obstacle on the vortex lattice structure and the associated dynamics of the condensate. In this scenario, the melting of the vortex lattice occurs upon either changing the position or oscillation frequency of the obstacle, and the threshold value for melting was found to be independent of the rotational frequency of the condensate. Additionally, we have observed that the angular momentum in the melted region exhibited the same power law exponent in the logarithmic scale regardless of the type of impurity. Further, the investigations related to the kinetic energy spectra revealed a similar power law behaviour in a melted state irrespective of the types of impurity. In conclusion, our analysis suggests different techniques to generate a turbulent state through vortex lattice melting in rotating BECs. These findings contribute to a deeper understanding of complex dynamics and structural transformations in rotating BECs.  

\section*{Acknowledgments}
R.B. acknowledges the financial support from the Department of Science and Technology's Innovation in Science Pursuit for Inspired Research (DST-INSPIRE) program, India. P.M. acknowledges the financial support from the MoE RUSA 2.0 program (Bharathidasan University - Physical Sciences).

\appendix
\counterwithin{figure}{section}
\section{Characterization of the structural transformation of vortices in rotating BECs with different rotational frequencies in the presence of random impurity potential}

\label{appendix}
In this appendix, we present a comprehensive analysis of the impact of various rotational frequencies on a condensate in the presence of random static impurities. %
\begin{figure}[!ht]
\centering
\includegraphics[width=0.95\linewidth]{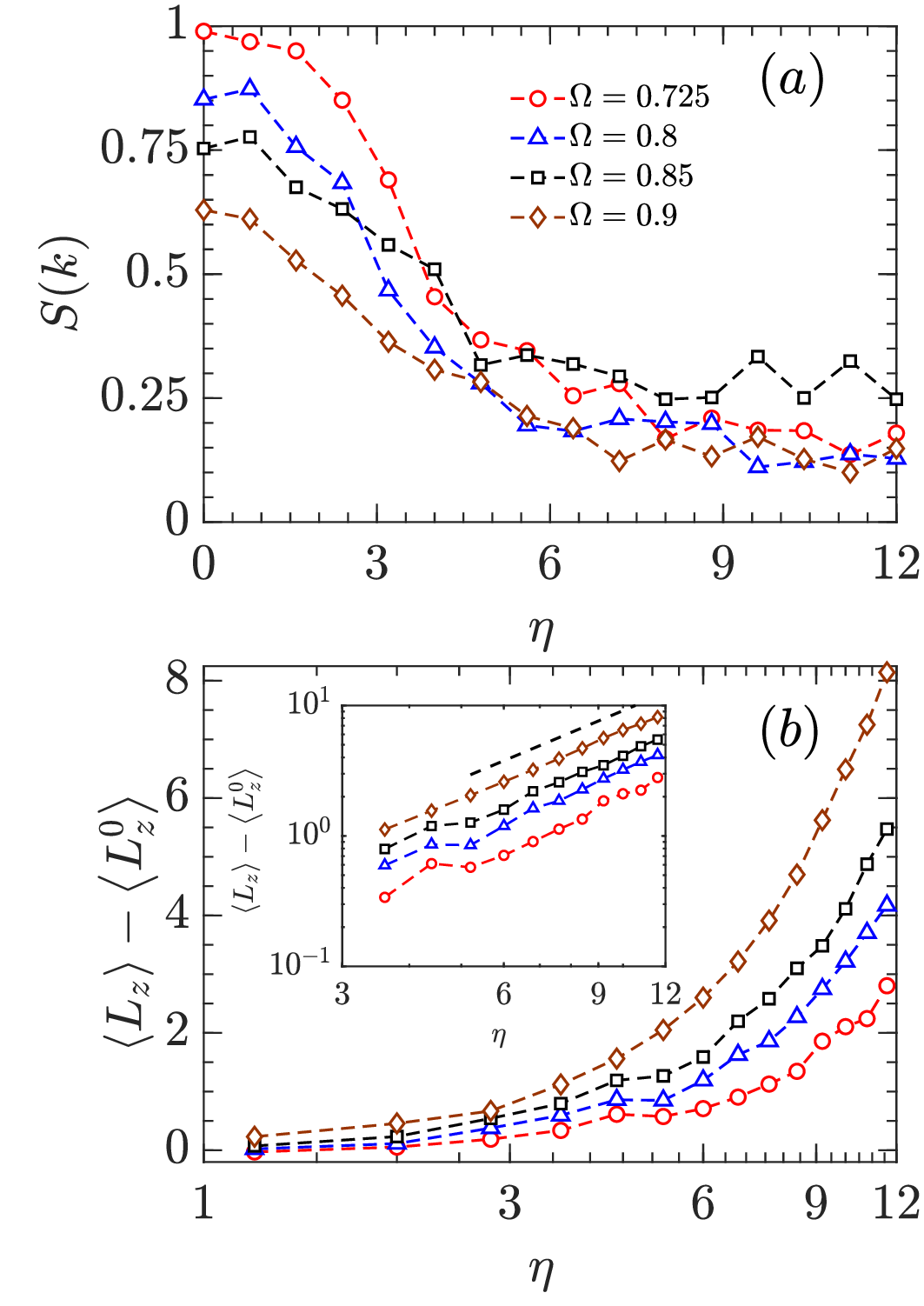}
\caption{Variation of (a) the structure factor, $S(k)$, and (b) the angular momentum, $\left\langle L_{z} \right\rangle  - \left\langle L^0_z\right\rangle$, as a function of $\eta$ for different $\Omega$ at $a_{\rm imp}=0.8$. Increasing $\Omega$ lowers the threshold value of $\eta$ above which the vortex lattice deforms, accompanied by a drop in $S(k)$ below $0.5$ in figure (a). The inset in (b) shows the variation of the same angular momentum on a log scale. The black dashed line in the log scale plot highlights the power-law behavior with respect to $\eta$. Here, $\left\langle L^0_z \right\rangle$ represents the angular momentum of the initial pinned vortex lattice.
}
\label{fig:ang_strct_diff_omg}
\end{figure}%

From  figure~\ref{fig:etac-php-static}, we have obtained that the critical value $\eta_c$ depends upon the rotational frequency $\Omega$ regardless of the impurity lattice constant $a_{\rm imp}$. Here, we present the corresponding structure factor and angular momentum to obtain the threshold value of $\eta_c$ for various $\Omega$ at $a_{\rm imp}=0.8$ as shown in figure~\ref{fig:ang_strct_diff_omg}. %
\begin{figure}[!ht]
\centering
\includegraphics[width=0.95\linewidth]{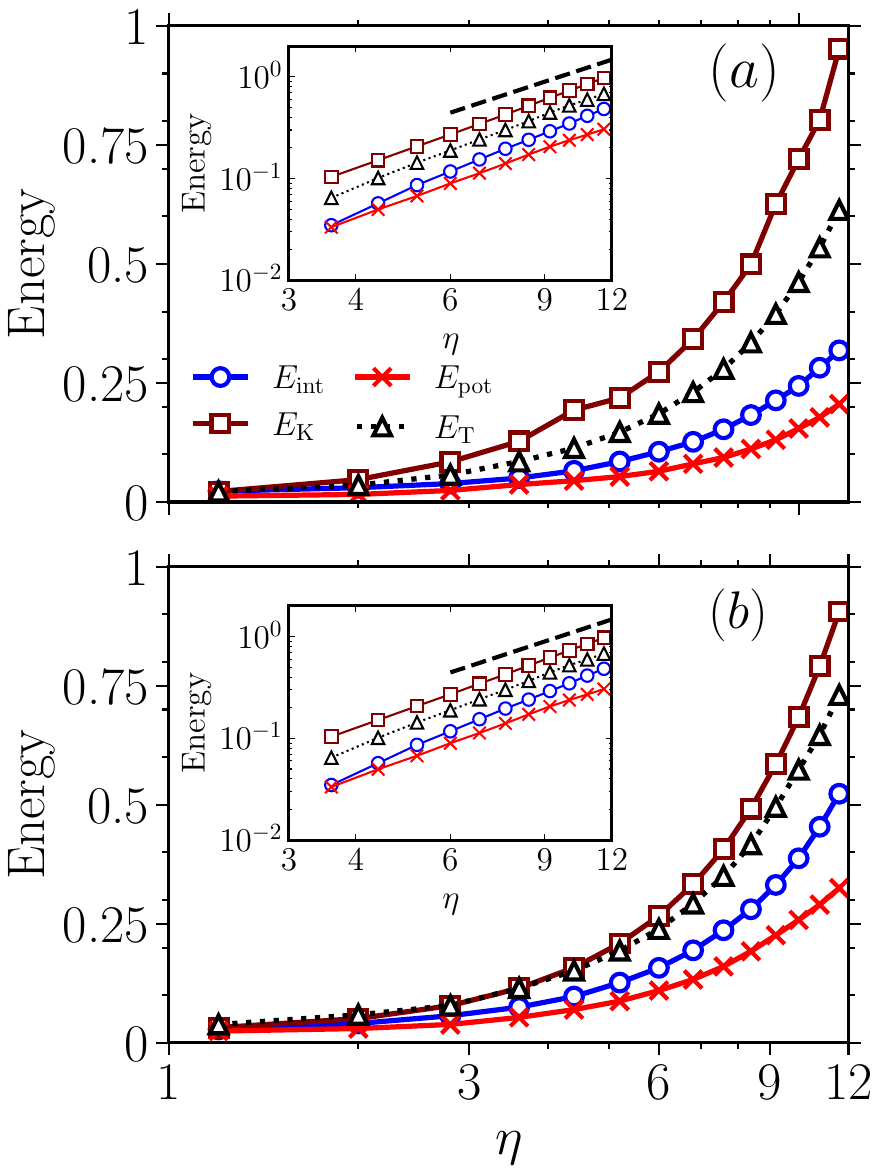}
\caption{Variation of different energies, $E_{int}$, $E_{pot}$, $E_{K}$, and $E_{T}$ as a function of $\eta$ for impurity lattice constant $a_{imp}=0.8$ for different $\Omega$: (a) $\Omega=0.725$, (b) $\Omega=0.9$. The inset shows the same in log scale. The black dashed line is drawn to show the power law behaviour of the energy components with in the melted state.}
\label{fig:energy_diff_omg}
\end{figure}%
We find that $S(k)$ gradually decreases upon increase in $\eta$ and the critical value  $\eta_c$ above which $S(k) < 0.5$ depends on $\Omega$. In figure~\ref{fig:ang_strct_diff_omg}(a), we illustrate the structural deformation of the vortex lattice. We find that the structural deformation of the vortex lattice occurs at $\eta=4$ and $\eta=3.2$ for $\Omega=0.725$ and $0.8$, respectively, however, for $\Omega=0.85$ and $\Omega=0.9$, it happens at $\eta=2.8$ and $2.4$ respectively. $\left\langle L_z \right\rangle $ increases upon increase in $\eta$, and the magnitude of $\left\langle L_z \right\rangle$ depends quite significantly on $\Omega$  as depicted in the figure \ref{fig:ang_strct_diff_omg}(b). The inset shows that, in the melting region, $\left\langle L_z \right\rangle $ has a power law dependence on $\eta$ with an exponent $1.73$. 

Next, in figure\ref{fig:energy_diff_omg}, we analyze the variation of kinetic~($E_{K}$), potential~($E_{pot}$), interaction~($E_{\rm int}$), and total energy~($E_{T}$) with respect to $\eta$ for different rotational frequencies: (a) $\Omega=0.725$, (b) $\Omega=0.9$ for $a_{\rm imp}=0.8$. We observe a consistent pattern in the behaviour of energy components, irrespective of the value of $\Omega$. Specifically, in the melted region, the kinetic energy~($E_K$) tends to dominate over the other energy components consistently, and the dominance of nonlinear energy over potential energy is observed across all frequencies. However, it is important to note that the rotational frequency ($\Omega$) affects the critical value at which the vortex lattice structure exhibits structural abnormalities. The inset of figure \ref{fig:energy_diff_omg}(a) presents a log-log plot of the energy components and reveals a power law dependence on $\eta$ with critical exponent $1.73$ in the melted region. 

\begin{figure}
\centering
\includegraphics[width=0.95\linewidth]{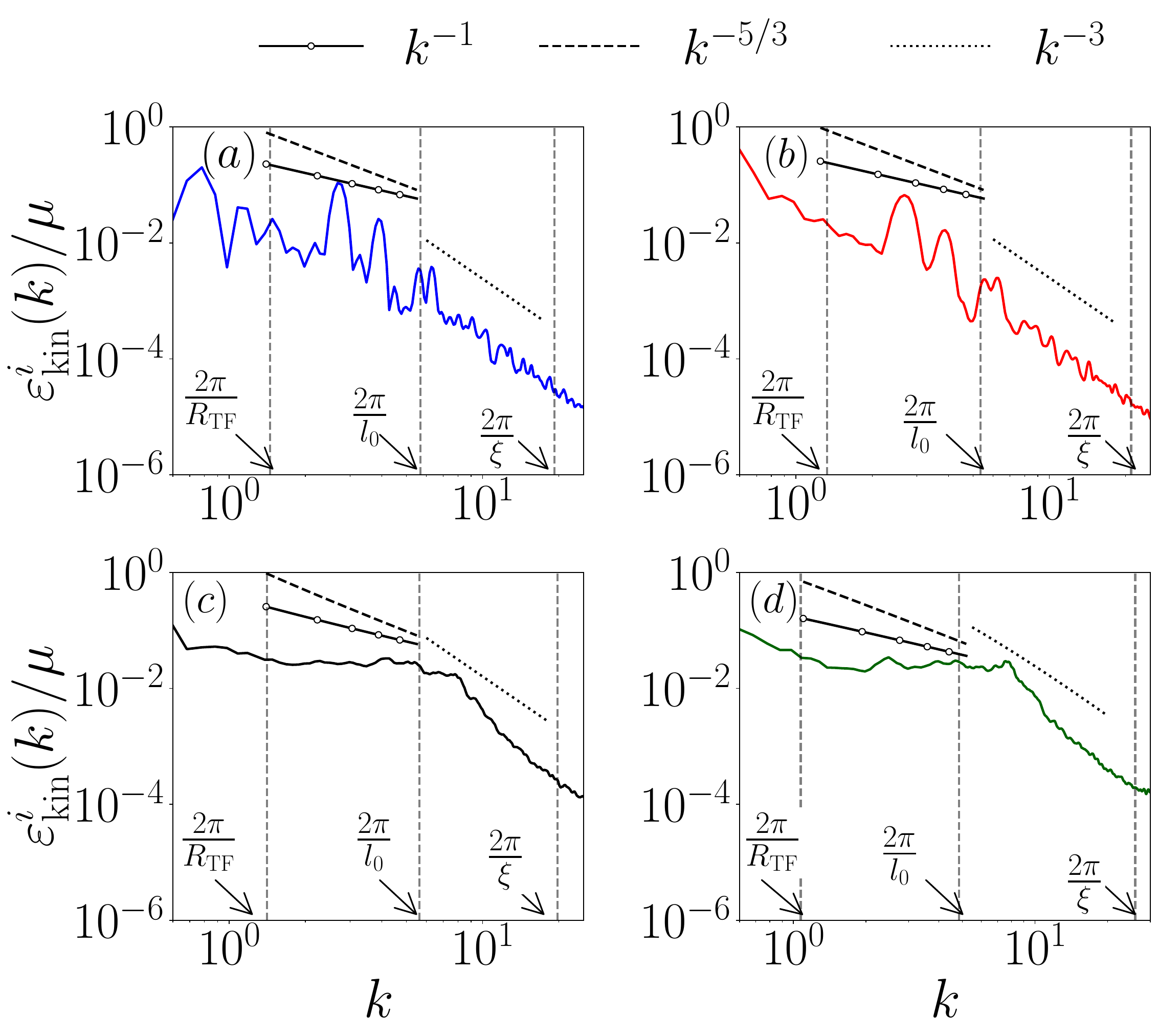}
\caption{Incompressible kinetic energy spectra for $\Omega=0.9$ for different $a_{\rm imp}$, and $\eta$:~(a) $a_{\rm imp}=2.2, \eta=3.6$; (b) $a_{\rm imp}=2.2, \eta=12$; (c) $a_{\rm imp}=0.8, \eta=3.6$; (d) $a_{\rm imp}=0.8, \eta=12$. It exhibits $k^{-1}$ and $k^{-3}$ power laws in the IR and UV regions, respectively.}
\label{fig:incomp_spectra:2}
\end{figure}

\begin{figure}[!ht]
\centering
\includegraphics[width=0.95\linewidth]{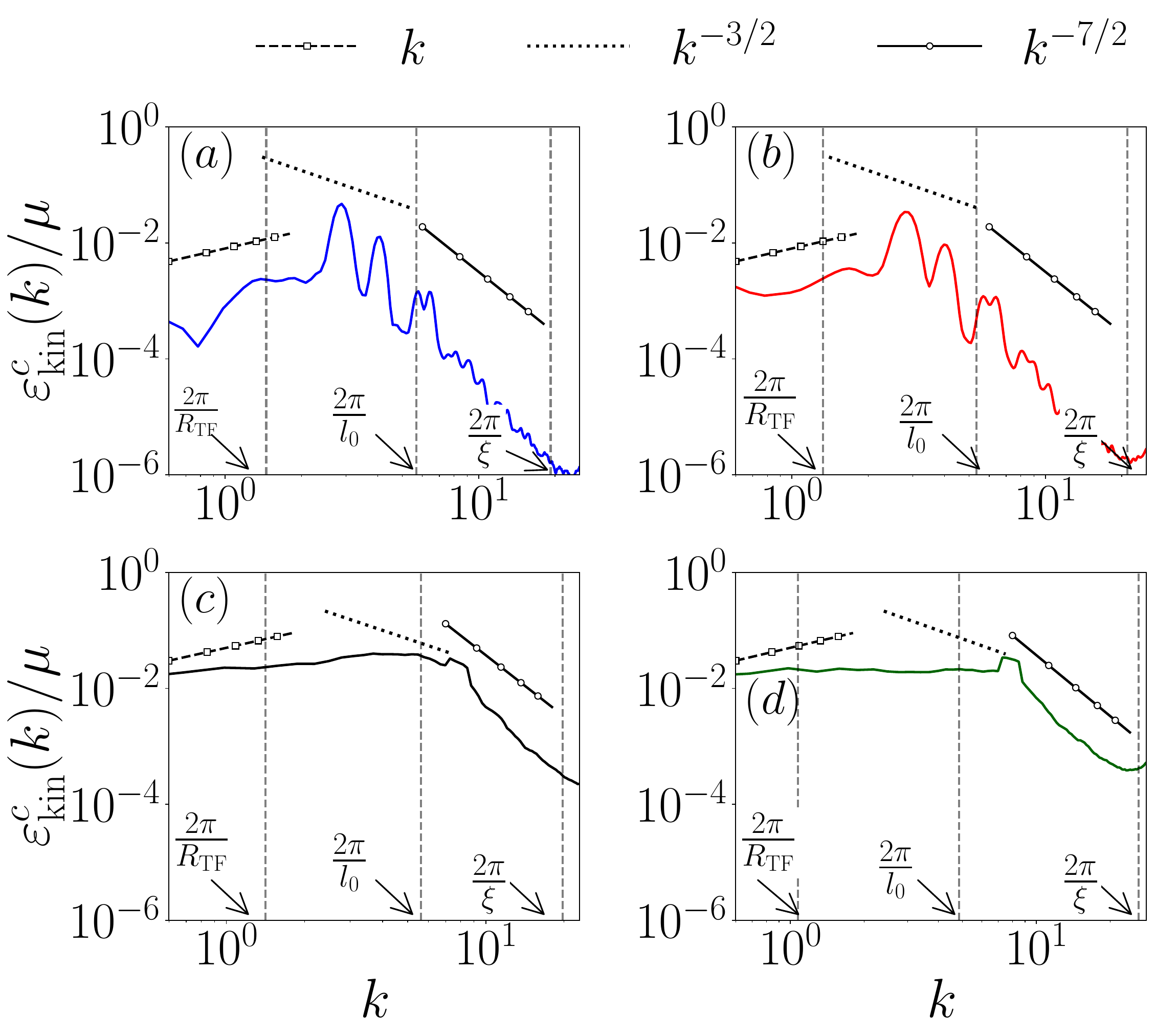}
\caption{Compressible kinetic energy spectra of the condensate at different impurity lattice strength ($\eta$) at $\omega=0.9$ for different $\eta$ and $a_{\rm imp}$:~(a) $a_{\rm imp}=2.2$, and $\eta=3.6$; (b) $a_{\rm imp}=2.2$, and $\eta=12$; (c) $a_{\rm imp}=0.8$, and $\eta=3.6$; (d) $a_{\rm imp}=0.8$, and $\eta=12$. The spectra display a $k$ power law for small value of $k$, while in the UV region it follows $k^{-3/2}$, and $k^{-7/2}$ power laws. }
\label{fig:comp_spectra:2}
\end{figure}
Figures~\ref{fig:incomp_spectra:2} and ~\ref{fig:comp_spectra:2} show the incompressible and compressible spectra corresponding to the melted vortex lattice state corresponding to the different $a_{imp}$ and $\eta$. In particular (a) for $a_{\rm imp}=2.2$, and $\eta=3.6$; (b) for $a_{\rm imp}=2.2$, and $\eta=12$; (c) for $a_{\rm imp}=0.8$, and $\eta=3.6$ and (d) $a_{\rm imp}=0.8$, and $\eta=12$. Like other spectrum bheaviour in the melted vortex lattice state the incompressible spectrum exhibits Vinen like spectrum ($\propto k^{-1}$) in the IR range of the wave number space for all the cases. However, the compressible spectra show the presence of the weak turbulence.

\providecommand{\newblock}{}


\begin{thebibliography}{10}
\expandafter\ifx\csname url\endcsname\relax
  \def\url#1{{\tt #1}}\fi
\expandafter\ifx\csname urlprefix\endcsname\relax\def\urlprefix{URL }\fi
\providecommand{\eprint}[2][]{\url{#2}}

\bibitem{Billy:2008}
Billy J, Josse V, Zuo Z, Bernard A, Hambrecht B, Lugan P, Cl{\'e}ment D, Sanchez-Palencia L, Bouyer P and Aspect A 2008 {\em Nature\/} {\bf 453} 891--894

\bibitem{Roati:2008}
Roati G, D’Errico C, Fallani L, Fattori M, Fort C, Zaccanti M, Modugno G, Modugno M and Inguscio M 2008 {\em Nature\/} {\bf 453} 895--898

\bibitem{Greiner2002}
Greiner M, Mandel O, Esslinger T, H{\"a}nsch T~W and Bloch I 2002 {\em nature\/} {\bf 415} 39--44

\bibitem{Leblanc2012}
LeBlanc L~J, Jim{\'e}nez-Garc{\'\i}a K, Williams R~A, Beeler M~C, Perry A~R, Phillips W~D and Spielman I~B 2012 {\em Proceedings of the National Academy of Sciences\/} {\bf 109} 10811--10814

\bibitem{Beeler2013}
Beeler M~C, Williams R~A, Jimenez-Garcia K, LeBlanc L~J, Perry A~R and Spielman I~B 2013 {\em Nature\/} {\bf 498} 201--204

\bibitem{Inouye:2001}
Inouye S, Gupta S, Rosenband T, Chikkatur A~P, G\"orlitz A, Gustavson T~L, Leanhardt A~E, Pritchard D~E and Ketterle W 2001 {\em Phys. Rev. Lett.\/} {\bf 87} 080402

\bibitem{Raman:2001}
Raman C, Abo-Shaeer J~R, Vogels J~M, Xu K and Ketterle W 2001 {\em Phys. Rev. Lett.\/} {\bf 87} 210402

\bibitem{Hodby:2001}
Hodby E, Hechenblaikner G, Hopkins S~A, Marag\`o O~M and Foot C~J 2001 {\em Phys. Rev. Lett.\/} {\bf 88} 010405

\bibitem{Williams1999}
Williams J and Holland M 1999 {\em Nature\/} {\bf 401} 568--572

\bibitem{Williams:2010}
Williams R~A, Al-Assam S and Foot C~J 2010 {\em Phys. Rev. Lett.\/} {\bf 104} 050404

\bibitem{Leanhardt:2002}
Leanhardt A~E, G\"orlitz A, Chikkatur A~P, Kielpinski D, Shin Y, Pritchard D~E and Ketterle W 2002 {\em Phys. Rev. Lett.\/} {\bf 89} 190403

\bibitem{Brachmann:2011}
Brachmann J~F~S, Bakr W~S, Gillen J, Peng A and Greiner M 2011 {\em Opt. Express\/} {\bf 19} 12984--12991

\bibitem{Madison:2000}
Madison K~W, Chevy F, Wohlleben W and Dalibard J 2000 {\em Phys. Rev. Lett.\/} {\bf 84} 806--809

\bibitem{Rosenstein:2010}
Rosenstein B and Li D 2010 {\em Rev. Mod. Phys.\/} {\bf 82} 109--168

\bibitem{Safar:1992}
Safar H, Gammel P~L, Huse D~A, Bishop D~J, Rice J~P and Ginsberg D~M 1992 {\em Phys. Rev. Lett.\/} {\bf 69} 824--827

\bibitem{Guillamon:2009}
Guillam{\'o}n I, Suderow H, Fern{\'a}ndez-Pacheco A, Ses{\'e} J, C{\'o}rdoba R, De~Teresa J, Ibarra M and Vieira S 2009 {\em Nat. Phys.\/} {\bf 5} 651--655

\bibitem{Engels:2003}
Engels P, Coddington I, Haljan P~C, Schweikhard V and Cornell E~A 2003 {\em Phys. Rev. Lett.\/} {\bf 90} 170405

\bibitem{Coddington:2003}
Coddington I, Engels P, Schweikhard V and Cornell E~A 2003 {\em Phys. Rev. Lett.\/} {\bf 91} 100402

\bibitem{Snoek:2006}
Snoek M and Stoof H~T~C 2006 {\em Phys. Rev. Lett.\/} {\bf 96} 230402

\bibitem{Mithun:2016}
Mithun T, Porsezian K and Dey B 2016 {\em Phys. Rev. A\/} {\bf 93} 013620

\bibitem{Mithun:2018}
Mithun T, Ganguli S~C, Raychaudhuri P and Dey B 2018 {\em Europhys. Lett.\/} {\bf 123} 20004

\bibitem{Hu:2020}
Hu P and Gu Q 2020 {\em J. Low Temp. Phys.\/} {\bf 199} 1314--1323

\bibitem{tsubota:2009}
Tsubota M 2009 {\em Contemp. Phys.\/} {\bf 50} 463--475

\bibitem{Henn:2009}
Henn E~A~L, Seman J~A, Roati G, Magalh\~aes K~M~F and Bagnato V~S 2009 {\em Phys. Rev. Lett.\/} {\bf 103} 045301

\bibitem{Gauthier:2019}
Gauthier G, Reeves M~T, Yu X, Bradley A~S, Baker M~A, Bell T~A, Rubinsztein-Dunlop H, Davis M~J and Neely T~W 2019 {\em Science\/} {\bf 364} 1264--1267

\bibitem{navon:2019}
Navon N, Eigen C, Zhang J, Lopes R, Gaunt A~L, Fujimoto K, Tsubota M, Smith R~P and Hadzibabic Z 2019 {\em Science\/} {\bf 366} 382--385

\bibitem{Kobayashi:2005}
Kobayashi M and Tsubota M 2005 {\em Phys. Rev. Lett.\/} {\bf 94} 065302

\bibitem{kobayashi:2007}
Kobayashi M and Tsubota M 2007 {\em Phys. Rev. A\/} {\bf 76} 045603

\bibitem{kobayashi:2008}
Kobayashi M and Tsubota M 2008 {\em J. Low Temp. Phys.\/} {\bf 150} 587--592

\bibitem{Kraichnan:1967}
Kraichnan R~H 1967 {\em Phys. Fluids\/} {\bf 10} 1417--1423

\bibitem{Kraichnan:1975}
Kraichnan R~H 1975 {\em J. Fluid Mech.\/} {\bf 67} 155--175

\bibitem{kolmogorov:1997}
Nore C, Abid M and Brachet M~E 1997 {\em Phys. Rev. Lett.\/} {\bf 78} 3896--3899

\bibitem{Reeves:2012}
Reeves M~T, Anderson B~P and Bradley A~S 2012 {\em Phys. Rev. A\/} {\bf 86} 053621

\bibitem{Neely:2010}
Neely T~W, Samson E~C, Bradley A~S, Davis M~J and Anderson B~P 2010 {\em Phys. Rev. Lett.\/} {\bf 104} 160401

\bibitem{Mithun:2021}
Mithun T, Kasamatsu K, Dey B and Kevrekidis P~G 2021 {\em Phys. Rev. A\/} {\bf 103} 023301

\bibitem{Subrata:2022}
Das S, Mukherjee K and Majumder S 2022 {\em Phys. Rev. A\/} {\bf 106} 023306

\bibitem{white:2012}
White A~C, Barenghi C~F and Proukakis N~P 2012 {\em Phys. Rev. A\/} {\bf 86} 013635

\bibitem{Sivakumar2024}
Sivakumar A, Mishra P~K, Hujeirat A~A and Muruganandam P 2024 {\em Phys. Fluids\/} {\bf 36} 027149

\bibitem{Sivakumar2024a}
Sivakumar A, Mishra P~K, Hujeirat A~A and Muruganandam P 2024 Dynamic instabilities and turbulence of merged rotating {Bose-Einstein} condensates (\textit{Preprint} \eprint{arXiv:2402.18474})

\bibitem{silva:2023}
da~Silva A~N, Kumar R~K, Bradley A~S and Tomio L 2023 {\em Phys. Rev. A\/} {\bf 107} 033314

\bibitem{Kasmatsu:2006}
Kasamatsu K and Tsubota M 2006 {\em Phys. Rev. Lett.\/} {\bf 97} 240404

\bibitem{Jin:1997}
Jin D~S, Matthews M~R, Ensher J~R, Wieman C~E and Cornell E~A 1997 {\em Phys. Rev. Lett.\/} {\bf 78} 764--767

\bibitem{Marago:2001}
Marag\`o O, Hechenblaikner G, Hodby E and Foot C 2001 {\em Phys. Rev. Lett.\/} {\bf 86} 3938--3941

\bibitem{Choi:1998}
Choi S, Morgan S~A and Burnett K 1998 {\em Phys. Rev. A\/} {\bf 57} 4057--4060

\bibitem{Kasamatsu:2002}
Tsubota M, Kasamatsu K and Ueda M 2002 {\em Phys. Rev. A\/} {\bf 65} 023603

\bibitem{Kasamatsu:2003}
Kasamatsu K, Tsubota M and Ueda M 2003 {\em Phys. Rev. A\/} {\bf 67} 033610

\bibitem{kasamatsu:2011}
Kato A, Nakano Y, Kasamatsu K and Matsui T 2011 {\em Phys. Rev. A\/} {\bf 84} 053623

\bibitem{Mithun:2014}
Mithun T, Porsezian K and Dey B 2014 {\em Phys. Rev. A\/} {\bf 89} 053625

\bibitem{Fort:2005}
Fort C, Fallani L, Guarrera V, Lye J~E, Modugno M, Wiersma D~S and Inguscio M 2005 {\em Phys. Rev. Lett.\/} {\bf 95} 170410

\bibitem{Sanchez:2010}
Sanchez-Palencia L and Lewenstein M 2010 {\em Nat. Phys.\/} {\bf 6} 87--95

\bibitem{Neely:2013}
Neely T~W, Bradley A~S, Samson E~C, Rooney S~J, Wright E~M, Law K~J~H, Carretero-Gonz\'alez R, Kevrekidis P~G, Davis M~J and Anderson B~P 2013 {\em Phys. Rev. Lett.\/} {\bf 111} 235301

\bibitem{Johnstone:2019}
Johnstone S~P, Groszek A~J, Starkey P~T, Billington C~J, Simula T~P and Helmerson K 2019 {\em Science\/} {\bf 364} 1267--1271

\bibitem{Kwon:2014}
Kwon W~J, Moon G, Choi J~y, Seo S~W and Shin Y~i 2014 {\em Phys. Rev. A\/} {\bf 90} 063627

\bibitem{Antoine:2014}
Antoine X and Duboscq R 2014 {\em Comput. Phys. Commun.\/} {\bf 185} 2969--2991

\bibitem{Madison:2001}
Madison K~W, Chevy F, Bretin V and Dalibard J 2001 {\em Phys. Rev. Lett.\/} {\bf 86} 4443--4446

\bibitem{Angela:2014}
White A~C, Anderson B~P and Bagnato V~S 2014 {\em PNAS\/} {\bf 111} 4719--4726

\bibitem{Madeira:2020}
Madeira L, Caracanhas M~A, dos Santos F and Bagnato V~S 2020 {\em Annu. Rev. Condens.\/} {\bf 11} 37--56

\bibitem{Bradley2022}
Bradley A~S, Kumar R~K, Pal S and Yu X 2022 {\em Phys. Rev. A\/} {\bf 106} 043322

\bibitem{Kumar:2019}
Kumar R~K, Chakrabarti B and Gammal A 2019 {\em J. Low Temp. Phys.\/} {\bf 194} 14--26

\bibitem{Pu:2005}
Pu H, Baksmaty L~O, Yi S and Bigelow N~P 2005 {\em Phys. Rev. Lett.\/} {\bf 94} 190401

\bibitem{li:2019}
Li X~L, Yang X~Y, Tang N, Song L, Zhou Z~K, Zhang J and Shi Y~R 2019 {\em New J. Phys.\/} {\bf 21} 103046

\bibitem{Marino:2021}
Marino {\'A}~V, Madeira L, Cidrim A, dos Santos F and Bagnato V~S 2021 {\em The European Physical Journal Special Topics\/} {\bf 230} 809--812

\end{thebibliography}
\end{document}